\documentclass[twocolumn, switch]{article} % Method A for two-column formatting
\usepackage{cite}
\usepackage{amsmath,amssymb,amsfonts}
\usepackage{algorithmic}
\usepackage{graphicx}
\usepackage{textcomp}
\usepackage[numbers]{natbib}
\usepackage{xcolor}
\usepackage[colorlinks=true, linkcolor=blue!50!black, citecolor=blue!50!black, urlcolor=blue!50!black]{hyperref}
\usepackage{etoolbox} % To prevent citation compression
\usepackage{float}
\usepackage{ragged2e}
\usepackage{amsmath}
\usepackage{amssymb}
\usepackage{mathrsfs}
\usepackage[utf8]{inputenc}

\usepackage{titlesec}
\usepackage[top=1.5cm, bottom=2.5cm, left=1.8cm, right=1.8cm]{geometry}
\usepackage{fancyhdr}
\usepackage{bm}
\makeatletter
\AtBeginDocument{\DeclareMathVersion{bold}
\SetSymbolFont{operators}{bold}{T1}{times}{b}{n}
\DeclareSymbolFont{NewLetters}{T1}{times}{b}{it}
\SetSymbolFont{NewLetters}{bold}{T1}{times}{b}{it}
\SetMathAlphabet{\mathrm}{bold}{T1}{times}{b}{n}
\SetMathAlphabet{\mathit}{bold}{T1}{times}{b}{it}
\SetMathAlphabet{\mathbf}{bold}{T1}{times}{b}{n}
\SetMathAlphabet{\mathtt}{bold}{OT1}{pcr}{b}{n}
\SetSymbolFont{symbols}{bold}{OMS}{cmsy}{b}{n}
\renewcommand\boldmath{\@nomath\boldmath\mathversion{bold}}}
\makeatother

\def\BibTeX{{\rm B\kern-.05em{\sc i\kern-.025em b}\kern-.08em
    T\kern-.1667em\lower.7ex\hbox{E}\kern-.125emX}}

\newcommand{\orcid}[1]{\href{https://orcid.org/#1}{\includegraphics[width=8pt]{Orcid_icon.png}}}

\usepackage{authblk}

\author{Nazmus Ashrafi}
\author{Salah Bouktif}
\author{Mohammed Mediani}

\affil[1]{Department of Computer Science and Software Engineering, United Arab Emirates University, Al Ain, United Arab Emirates}

\title{Enhancing LLM Code Generation: A Systematic Evaluation of Multi-Agent Collaboration and Runtime Debugging for Improved Accuracy, Reliability, and Latency}
\begin{document}

\twocolumn[ % Method A for two-column formatting
  \begin{@twocolumnfalse} % Method A for two-column formatting

\maketitle

\begin{abstract}
\vspace{0.35cm}
The use of large language models (LLMs) for automated code generation has emerged as a significant focus within AI research. As these pretrained models continue to evolve, their ability to understand and generate complex code structures has opened new possibilities for automating intricate programming tasks for the sake of accurate code generation. Although contemporary foundational models demonstrate promoting results, researchers continue to explore optimal post-training strategies to enhance code quality. These include supervised fine-tuning, retrieval-augmented generation (RAG), debugging, and many others. In this paper, we combine  two widely used approaches namely (1) multi-agent collaboration and (2) runtime execution information-based debugging— for improving code generation functionality, reliability, and practical applicability. We perform an  empirical study in order to extend the evaluation of the individual strategies as well as the proposed composition of the activities of both strategies. Our study use 19 LLMs to examines the performance of individual and the proposed strategies, offering comprehensive insights into how different programming activities compositions and training paradigms influence code generation effectiveness. In particular, we implement a chained system that combines both strategies to assess their combined impact on functional accuracy, code reliability, and generation latency using two benchmark datasets commonly used for code generation. Our findings provide valuable insights for organizations seeking robust AI-driven coding solutions by guiding them in selecting models that can better adapt to complex post-training strategies, ultimately fostering the adoption of more effective and reliable code generation technologies.
\end{abstract}
%\keywords{First keyword \and Second keyword \and More} % (optional)
\vspace{0.35cm}

  \end{@twocolumnfalse} % Method A for two-column formatting
] % Method A for two-column formatting

%\begin{multicols}{2} % Method B for two-column formatting (doesn't play well with line numbers), comment out if using method A

\maketitle

\section{Introduction}
\label{sec:introduction}
Large Language Models (LLMs) have significantly transformed the field of Natural Language Processing (NLP), driving advancements in a wide range of tasks, including chatbots, text summarization, and content generation. Recently, numerous AI research laboratories have focused on developing increasingly sophisticated models that push the boundaries of what these systems can achieve. Among the most promising applications of LLMs is code generation, a key area that has gained significant attention due to its potential to automate complex programming tasks \citep{13,14,15}. To enhance the quality and reliability of code generated by LLMs, many post-training strategies have been proposed. Post-training strategies refine foundational models for improved task performance through techniques such as supervised fine-tuning, direct preference optimization, and retrieval-augmented generation, among others. Certain post-training strategies can be applied without the need for the costly fine-tuning process, which involves updating all model parameters with new training data. These strategies include techniques such as retrieval-augmented generation (RAG) and multi-agent conversational feedback-based generation, among others. Furthermore, certain post-training strategies are tailored for tasks similar to code generation, such as leveraging program execution information for iterative debugging to enhance code quality and generating additional test cases to improve the robustness of the generated code. These approaches can improve the performance of weaker models while maintaining cost efficiency. The underlying logic and architecture of these strategies play a crucial role in determining model effectiveness for the assigned task. In this paper, we conduct a comprehensive analysis of two key post-training approaches—multi-agent collaboration and runtime debugging—by evaluating their impact across a diverse set of models. Furthermore, we implement a unified framework that integrates elements of multi-agent collaboration with runtime debugging to enhance LLM-generated code. Our study identifies the model characteristics that most effectively respond to this combination, offering valuable insights into the conditions that optimize their impact.

The multi-agent collaboration strategy \citep{16, 17, 18, 19, 20, 21} leverages the interaction of multiple LLM-empowered agents, each contributing to different aspects of the code generation process. By establishing dynamic role-playing interactions between agents, this method creates a cooperative environment that mirrors the division of labor typically practiced by software development teams. For example, a plan generated by a program analyst agent can guide a coder agent in producing more effective and optimized code. The idea is that by assigning each agent a specialized implementation task, their collaboration enhances the accuracy and efficiency of the code generation process. The second strategy is based on the use of runtime execution feedback to perform better informed debugging \citep{22, 23, 24, 25, 26, 27}. The feedback captured provides the LLM with insight into the localized successes and failures of the early versions of the generated code. In other terms, by assessing the correctness of the execution flow of code segments, this approach provides valuable context about potential issues and the overall flow of the program. By integrating rich context, LLMs can improve their code generation over time, ensuring better reliability and functional correctness. Both of these post-training techniques represent promising avenues for enhancing LLM-driven code generation, offering new opportunities for improving pre-trained model performance and usefulness in real-world programming scenarios.  

Software engineering encompasses two fundamental viewpoints, namely, the process-oriented and the product-oriented points of view, each influencing how software is developed and evaluated. The process-oriented viewpoint emphasizes structured methodologies, best practices, and workflow efficiency, focusing on how software is built. It is based on the premise that a well-defined and rigorous process leads to high-quality results. In contrast, the product-oriented viewpoint prioritizes the quality and effectiveness of the final product, emphasizing its usability, relevance, and continuous improvement through iterative evaluation, rapid prototyping, and experimentation. Although these philosophies represent distinct perspectives, the principles underlying them are complementary, offering a balanced approach to software development. This synergy between the two philosophies motivates their integration into a combined approach, as outlined in the following paragraph.

In this work, we implement a framework that combines the two prominent post-training strategies discussed above—multi-agent collaboration as a process-oriented pipeline to generate the code and runtime execution information-based debugging as a product-oriented evaluation mechanism—into a chained system. This combined approach, which begins with multi-agent collaboration and culminates in debugging, mirrors a real-world software development workflow. In such a scenario, a process-oriented team, consisting of an analyst, coder, and tester, collaborates to produce the code. This collaborative effort represents the multi-agent collaboration process, where each agent contributes their expertise to ensure the quality and functionality of the code. Once the code is generated, it is compiled and run to identify potential issues. If errors are detected, the generated code (the product) is debugged using runtime execution information-based debugging, where breakpoints are strategically placed to capture execution details, thereby enabling the refinement of the product. The feedback obtained during this debugging process informs the coder, helping to iteratively improve the code. 

To evaluate the impact of this combined methodology, we examine various performance metrics such as functional accuracy, code rigor, and generation latency. Specifically, we compare the performance of three segments of our chained system, as well as the complete chained system, against the original approaches: multi-agent collaboration and debugging. This comparison enables us to assess how the combinations of these strategies enhance code generation compared to each strategy individually, providing a comprehensive analysis of their potential benefits. 

Our experiments are comprehensive, covering 19 LLMs with variations in size, training data mixture, and training methodologies. Using two benchmark datasets and five baseline approaches, these experiments reveal several key insights into the performance of the original techniques and their combinations. First, debugging-based approaches generally outperform agentic workflows in most cases, at least within the specific methods we examined. Through empirical evaluation, we were able to statistically demonstrate, with a confidence level of 85\% (alpha = 0.15), that the combined approach offers superior performance compared to the agentic workflow alone. However, we could not establish a statistically significant improvement when comparing the combined approach with the debugging technique in isolation. Interestingly, the combination of a straightforward agentic workflow, such as a two agent Analyst-Coder collaboration, with the debugging mechanism results in a slight improvement in mean accuracy (0.68\%) across the 19 LLMs and two datasets compared to debugging alone, but adding a more complex agentic workflow—such as a three agent Analyst-Coder-Tester system—reduces accuracy for most models. 

Further observations reveal that when the performance gap between the debugging and agentic approach is relatively small for a particular LLM, combining the two methods tends to produce better outcomes. However, when the debugging approach significantly outperforms the agentic one, adding the latter may introduce unnecessary complexity, diminishing the overall benefit. Our observations also indicate that while simple agentic workflows, such as the Analyst-Coder system, are capable of producing more rigorous code, increasing the complexity of agentic interactions often leads to less performant outputs. Based on these findings, we determine that combining a straightforward agentic workflow, such as the Analyst-Coder system, with the debugging mechanism achieves an optimal balance. This approach not only improves functional accuracy across a range of LLMs but also delivers rigorous code with latency comparable to that of the standalone debugging technique. Furthermore, we observe that OpenAI models generally perform well when combination approaches are applied. These findings emphasize the potential of combining complementary post-training strategies to improve the reliability and efficiency of LLM-driven code in widely used AI products, particularly those performing downstream tasks like code generation, where even a marginal improvement in functional accuracy can have a substantial impact.

The remainder of this paper is structured as follows. Section II presents the literature review, followed by a detailed description of our research methodology in Section III. In Section IV, we outline the research questions and the experimental setup, including the chosen metrics, datasets, and baselines. We then discuss our experimental results in relation to the research questions. Finally, Section V concludes the study and highlights the key findings.

\section{Literature Review}
This section presents a theoretical literature review, examining two post-training methods commonly used to enhance the code generation capabilities of pretrained LLMs. These widely adopted approaches in the field are—multi-agent collaboration and debugging based on program runtime execution information.

\subsection{Multi-agent collaboration}

Multi-agent frameworks powered by LLMs have attracted considerable interest from both academic and industrial sectors. This approach employs intricate agentic interactions to establish a robust foundation for code generation. For example, AgentCoder \citep{16} is a multi-agent framework for code generation where a LLM powered test designer agent creates test cases, and a test executor agent runs the code, providing feedback to the coder agent for refinement. MapCoder \citep{17} is a multi-agent system designed to replicate the human programming process utilizing four LLM powered agents: Retrieval, Planning, Coding, and Debugging. These agents operate in a sequence, with each stage building upon the previous one through dynamic interaction and feedback, improving code generation by leveraging relevant examples,  structured planning, and cross-checking. The self-collaboration framework in \citep{21} employs three LLM agents in a structured workflow: the Analyst Agent decomposes coding tasks and creates a high-level plan, the Coder Agent implements the solution based on this plan, and the Tester Agent iteratively evaluates and guides the Coder Agent to refine the code. The RGD framework \citep{25} utilizes a multi-LLM debugging approach with three agents: the Guide Agent performs initial reasoning and retrieves guidance from a memory pool, the Debug Agent generates code based on this guidance, and the Feedback Agent evaluates execution outcomes to deliver failure analysis, enhancing future iterations. FlowGen \citep{28} is a multi-agent framework for code generation that simulates software process models (Waterfall, Test-Driven-Development, Scrum) through LLM agents assigned to various development roles. It leverages organized interactions, iterative refinement, and continuous feedback to improve the quality of generated code. Multi-agentic interaction offers a promising avenue for enhancing code generation with LLMs, as it facilitates role-playing-based collaboration and division of labor—key attributes of an effective software engineering team in real-world scenarios.

\subsection{Runtime execution information-based debugging}

Leveraging runtime execution information to enhance code generation is a widely recognized approach in the field, with many frameworks employing this method garnering considerable attention. For instance, Cycle \citep{22} allows pretrained LLMs to self-refine by continuing training with data samples derived from three information sources: high-level problem descriptions, incorrect code generated previously, and execution feedback from the executor when the code fails, leading to improved code generation. \citep{24} introduces a print debugging method in which the LLM inserts 'print statements' or 'log messages' into the code to capture insights into the execution flow and variable values, which are then used in subsequent iterations to improve code generation. LDB \citep{23} improves debugging by performing static analysis to construct a control flow graph, then using test cases and execution traces to monitor intermediate variables in each basic block. It queries an LLM to assess the correctness of each block and explain the execution flow, using this feedback to refine the code. ReflectionCoder \citep{29} generates reflection sequences that merge compiler feedback with the model’s analysis of execution errors. These sequences are used to train models, enabling them to enhance their code generation performance through learning from iterative corrections. MGDebugger \citep{26} is a hierarchical debugging tool that addresses bugs by breaking code down into sub-functions of varying granularity. It uses an LLM-simulated Python executor to monitor code execution and track variable states, resolving issues through a bottom-up approach. The self-debugging framework \citep{30} leverages LLM to iteratively debug code through generation, explanation, and runtime execution feedback, enabling models to identify and correct errors without human intervention. These works have significantly advanced code generation tasks by leveraging runtime execution-based information for debugging, effectively demonstrating its value in enhancing code generation with LLMs.

\section{Methodology}

Our research uses an integrated approach that combines structured collaboration in code generation with a targeted debugging mechanism. The initial phase adopts a process-oriented strategy, where agents with specialized roles—Analyst, Coder, and Tester—work together to facilitate an organized and methodical code generation process. This multi-agent collaboration reflects established software development workflows, emphasizing specialization and adherence to best practices to produce high-quality code. Following this, the generated code undergoes a product-oriented evaluation through a structured debugging phase, utilizing runtime execution data to identify and resolve issues effectively. By linking collaborative development with targeted debugging, this framework leverages the strengths of both process-driven methodologies and product-focused refinement, ensuring a balanced and effective approach to LLM-assisted coding. The complete pipeline is depicted in \hyperref[figure1]{Figure 1}.

As shown in \hyperref[figure1]{Figure 1}, the framework functions in two main phases. The first phase employs a multi-agent collaboration system in which three specialized LLM agents work in conjunction following a modified waterfall model \citep{21}. The second phase incorporates runtime execution information for debugging, which is activated when initial code generation requires refinement. We selected this chain alignment due to its natural progression, mirroring the typical workflow of a software engineer: planning, writing code, and debugging block-by-block when errors are detected during execution.

We begin by segmenting the HumanEval benchmarking dataset into three components: the task description, which encompasses code snippets and natural language requirements; the visible test cases, employed to execute the code within the framework; and the hidden test cases, reserved for evaluating the final output. The framework implements a sequential flow in which the code is first generated using the ACT (analyst, coder, and tester agents) collaboration system. This phase continues until the generated code successfully passes execution against the visible test cases or the maximum allowed iterations (retriesCT) of coder-tester interactions are completed. A limitation of this approach is its reliance on a limited set of visible test cases, which may not capture all edge cases, leaving some potential errors undetected. When problems are detected or the limit is reached, the system activates the debugging phase, where the code that failed the test is analyzed and refined using runtime execution information. This process continues until either a successful solution is achieved or the maximum number of debugging iterations (retriesD) is reached. If the maximum number of debugging iterations is reached, and the code is still unable to produce a correct result, the latest version of the code is deemed final. Agents utilized in both phases are provided with role-specific instructions and general guidelines (see our GitHub repository\footnote{\label{fn:github}\url{https://github.com/nazmus-ashrafi/multiagent_vs_debugger}} for prompts). We used the same set of prompts for all models, which may not be ideal but was necessary to maintain consistency throughout the experiment.

The combination of these two approaches creates a robust system that benefits from collaborative problem-solving and systematic debugging. The ACT system delivers high-level architectural guidance and evaluation to facilitate the generation of quality code aligning with process-oriented principles, while the debugging phase provides detailed, execution-driven refinement of the code produced by the ACT module, adhering to a product-oriented approach that emphasizes strong evaluation measures. However, a potential caveat to this combined approach is that if the agentic conversation in the ACT phase results in extremely poor initial code generation, it could complicate the debugging process. Flawed initial code can misdirect the debugging phase, emphasizing the need for strong ACT performance by a model to ensure the success of our integrated technique.

\subsection{Multi-Agent Collaboration Phase}

In the initial phase of our combined approach, which represents the process-oriented part of our framework, we implement a multi-agent collaboration system involving an Analyst, Coder, and Tester, similar to the self-collaboration framework described in \citep{21}. This configuration was selected because of its simplicity and the manageable number of agents, which makes it a suitable representation of typical agent-based workflows in this domain. Additionally, it offers flexibility, as the complexity of the agentic interactions can be easily adjusted by removing the agents from the mix.

Building on this foundation, our framework facilitates collaboration among three distinct LLM agents within a structured workflow. The Analyst Agent begins by decomposing the coding requirements into manageable subtasks and developing a high-level implementation plan. This abstraction process aims to reduce complexity and provide clear guidance for the subsequent coding stage. The Coder Agent then implements the solution following the analyst's architectural blueprint. The implementation undergoes iterative refinement through interaction with the Tester Agent, which evaluates the functionality, readability, and maintainability of the code \citep{21}. This interaction can occur up to three times (retriesCT = 3), a modification from the original framework's four iterations, optimized to balance thoroughness with computational efficiency.

\subsection{Debugging Phase}

In the final phase of our combined approach, we incorporate a system similar to the Large Language Model Debugger (LDB) \citep{23} as the last component of our chained framework, which represents the product refinement aspect. LDB was chosen for its simplicity and its effective portrayal of debugging practices in real-world software engineering environments, though it does not fully replicate the exact workflow encountered in practice. Addressing this gap is left as a direction for our future work. Furthermore, combining LDB with Reflexion \citep{44} has been shown to achieve state-of-the-art results, achieving a remarkable 98.2 score on the HumanEval dataset using OpenAI’s GPT-4o. At the time of our experiments, LDB held the top position on the leaderboard\footnote{\label{fn:leaderboard}\url{https://paperswithcode.com/sota/code-generation-on-humaneval}}.  This exceptional performance highlights the potential of integrating LDB into agentic workflows, making it a compelling subject for deeper analysis, particularly in understanding how it adapts to varying levels of complexity in multi-agent collaboration across a diverse range of LLMs.

The framework transitions to the debugging phase when the code generated by the ACT system does not pass a set of visible test cases during execution. This phase employs a debugging mechanism that decomposes the code that failed the test into basic blocks using a Control Flow Graph (CFG) analysis. The LDB paper evaluates program decomposition at three levels: line-level, block-level , and function-level. We adopt the block-level implementation because it achieved the highest accuracy. During runtime execution with the visible test case that caused the initial failure, intermediate variable values within these blocks are monitored and logged, capturing detailed contextual information about the failing program’s behavior. This information is provided to a debugger agent whose role is to analyze the execution of each block and identify potential errors, offering explanations for why a particular block may be incorrect. These explanations are subsequently passed to a coder agent, who utilizes them to generate the corrected code. By integrating execution-level insights into the coding process, this phase ensures that issues are identified and resolved systematically, enhancing the overall quality of the generated code. The debugging cycle can be repeated up to four times (retriesD = 4), a practical limitation implemented to optimize resource utilization while maintaining effective error resolution. This represents a strategic reduction from the original LDB framework's ten-iteration maximum, accounting for the increased computational demands of our combined approach. 

\begin{figure*}[!htb]
    \centering
    \includegraphics[width=\textwidth,clip,keepaspectratio]{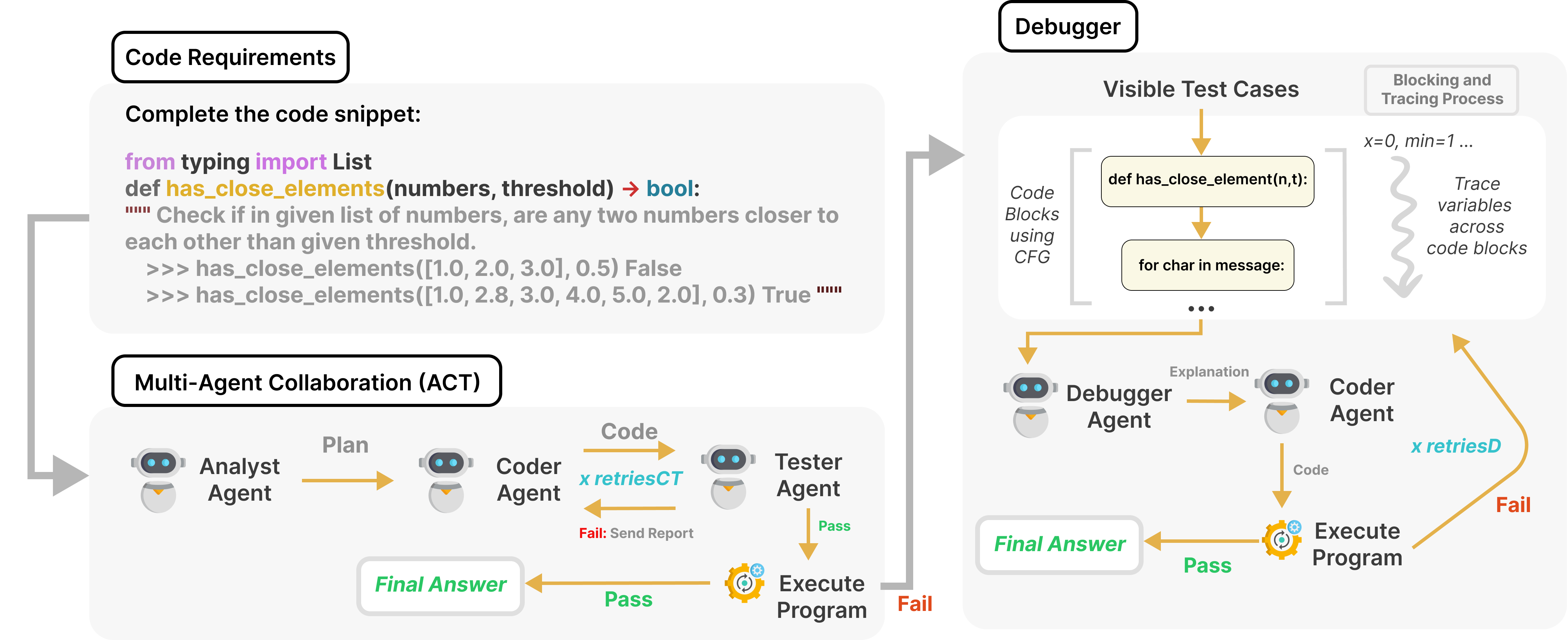}
    \caption{\centering Architecture of ACT + Debugger approach.}
    \label{figure1}
\end{figure*}

\section{Evaluation}
\vspace{0.5cm}
\subsection{Experiment Setup}

Our goal is to address the following research questions (RQs):

\begin{itemize}
\item RQ1: How does combining two widely adopted post-training techniques for enhancing LLM code generation—multi-agent collaboration and run-time execution information-based debugging—impact functional accuracy compared to applying each technique individually?

\item RQ2: How does each segment of our combined system influence the functional accuracy of the generated code, and which segmented approach achieves the highest functional accuracy when considered independently?

\item RQ3: How rigorous is the code generated by each segmented approach, particularly in maintaining accuracy under more stringent testing conditions, and which models excel in producing the most rigorous code?

\item RQ4: How does the latency of each segmented approach compare and what are the trade-offs between execution time and the quality of the generated code?
\end{itemize}

\subsubsection{Metric}
{We use the pass@k metric to judge the functional accuracy of the generated code. The Pass@k metric is widely used to evaluate the performance of code generation models \citep{14, 15, 31}. Originally proposed in \citep{13}, the pass@k metric measures the probability that at least one of the top generated samples for a problem is correct. Pass@k provides a practical assessment of the model’s ability to produce functionally correct code within its top attempts.

To simplify the experimental setup, we chose to generate only one sample per problem (n=1) in our experiments. This eliminates the complexity of evaluating multiple samples and focuses entirely on pass@1 as the evaluation metric. When only one sample is generated, pass@k effectively reduces to a binary outcome: either the single sample is correct (pass@1 = 1) or it is not (pass@1 = 0). With only one sample per problem, the evaluation directly reflects how reliably the model can generate correct solutions without additional attempts, simulating practical use cases where developers rely heavily on the first suggestion provided by the model.
}

\subsubsection{Datasets}

In this experiment, we evaluate our proposed chain and the individual techniques using two extensively adopted code generation datasets, HumanEval \citep{13} and HumanEval+ \citep{32}.

HumanEval was originally proposed by OpenAI and consists of 164 handwritten programming tasks. Each task consists of a function signature, a natural language (NL) description, example use cases, the function’s implementation, and multiple unit tests (with an average of 7.7 tests per task). HumanEval encompasses a variety of programming challenges, designed to assess the model’s problem-solving abilities and flexibility across a broad spectrum of problems. It is commonly used to assess the coding capabilities of LLMs \citep{1, 2, 3, 9, 33, 34, 35, 36, 37}.

HumanEval+ enhances the original HumanEval dataset with 80 times more tests. Examining the score variations between HumanEval and HumanEval+, particularly before and after applying the HumanEval+ tests, reveals key insights. A smaller decline indicates more rigorous code generation, while a larger drop suggests that the generated code is more prone to fragility. It is a widely recognized and rigorous benchmark, and many LLMs \citep{1, 2, 3, 38, 39, 40, 41, 42} consistently appear on its leaderboard.

\begin{table*}[!htbp]
    \centering
    \includegraphics[width=10.8\textwidth,height=8.0in,clip,keepaspectratio]{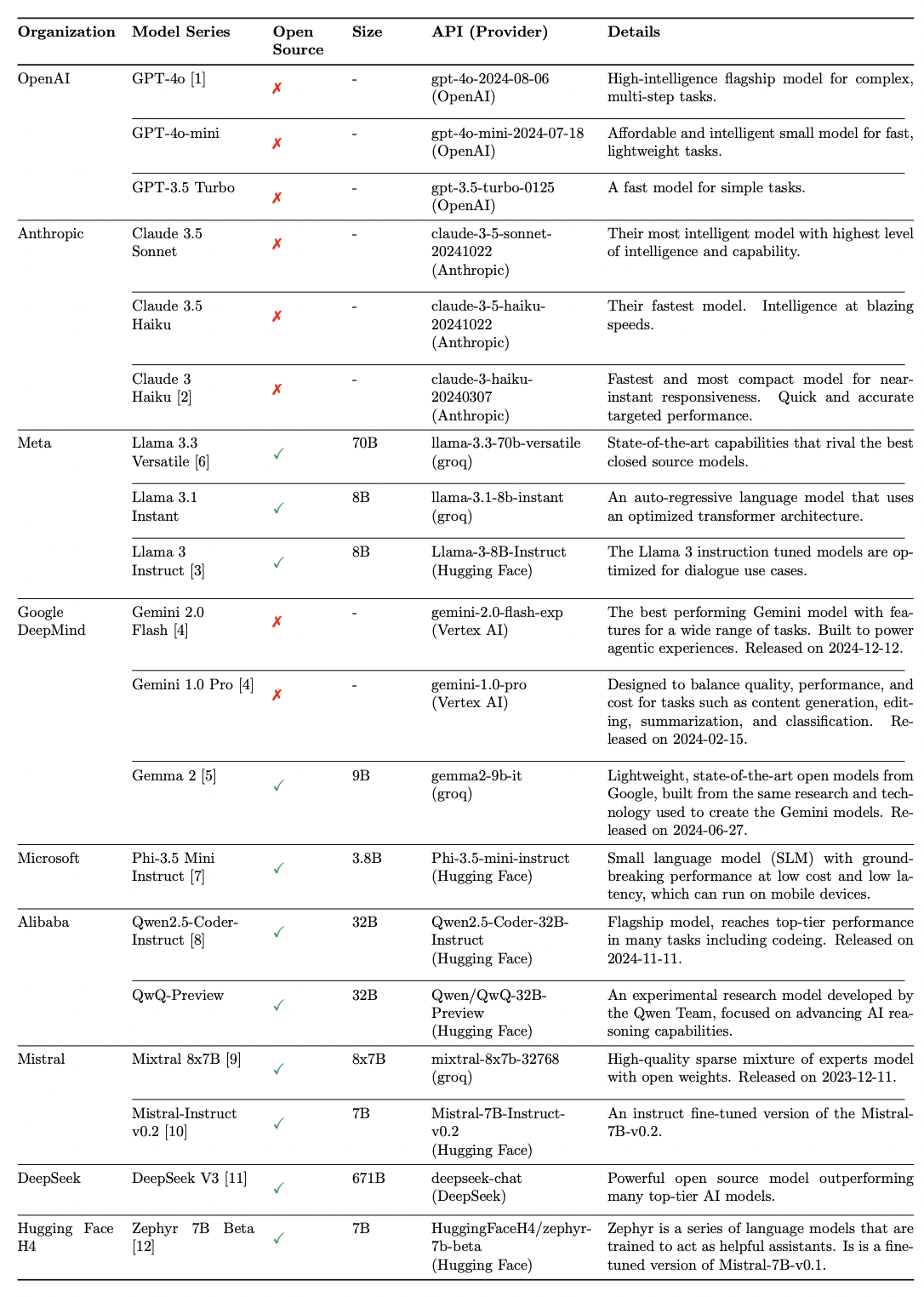} % Adjust the width as needed
    % \caption{Information table for LLMs used in this work}
    \caption{\centering Table summarizing the LLMs utilized in this study (All APIs were accessed in the month of December 2024).}
    \label{table1}
\end{table*}

\subsubsection{Baseline}

To evaluate the performance of our proposed combined approach (ACT + Debugger), we compared it against its individual components and configurations (segments), each representing a distinct approach to code generation. These approaches are as follows:

\begin{enumerate}
    \item Basic: In this method, we simply prompt the LLM with the problem description, and the model generates code without any additional refinement or collaboration.
    \item AC (Analyst and Coder Collaboration): Here, the analyst agent generates a plan which is then followed by the coder agent to produce the code. This approach involves a single interaction between the analyst and the coder.
    \item ACT (Agent-Coder-Tester Collaboration): This method incorporates three agents: the analyst provides a plan, the coder generates code, and the tester reviews the generated code, offering feedback for refinement. The coder and the tester can iterate up to three times to improve the code.
    \item Debugger Only: This technique focuses on debugging. The code is divided into basic blocks, and the values of intermediate variables are tracked throughout runtime execution. The debugger ensures the correctness by verifying each block against the task description.
    \item AC + Debugger: After the analyst and coder agents collaborate to generate code, the resulting code is subjected to debugging to identify and correct errors through runtime execution information.
\end{enumerate}

Additional details on the composition of these approaches are provided in the RQ2 section, where approach-wise performance is discussed. These approaches serve as a baseline or reference points to assess the effectiveness of ACT + Debugger, which incorporates both multi-agent collaboration and debugging within a unified approach, as illustrated in \hyperref[figure1]{Figure 1}. ACT and Debugger are presented as separate modules in \hyperref[figure1]{Figure 1} to illustrate how these baseline techniques can be independently structured. By comparing the baseline approaches, we aim to demonstrate how combining agent collaboration with debugging leads to more rigorous and accurate code generation.

\subsubsection{Models}
We evaluated the baselines and our proposed approach on 19 models with variations in size, training data composition, context length, and training techniques. Key information about these models and the API endpoints that were used to access the models are shown in \hyperref[table1]{Table 1}. Although evaluating every LLM on these tasks is impractical, we strive to provide a comprehensive assessment of the baselines and the proposed approach by exploring a diverse range of models of varying sizes, trained data mixtures, and training styles, spanning both open and closed-source domains. We include powerful generalist models like OpenAI’s GPT-4o, instruction trained models like Meta’s Llama 3 Instruct, and a code specific model (Qwen 2.5 Coder Instruct from Alibaba). We test open source models range from 3.8B parameters (e.g. Microsoft’s Phi 3.5 Mini) which is meant to run on portable devices to 671B parameters (e.g. DeepSeek AI’s DeepSeek-V3). We also include closed-source LLMs, presumed to have over 175 billion parameters (e.g. OpenAI’s GPT-4o). 

We integrate leading models from major AI research organizations that encompass both open-source and closed-source architectures available at the time of this study. This selection ensures that our assessment provides a comprehensive representation of model performance across a diverse range of cost and size brackets. The diverse choice of LLMs enhances the generalizability of our findings and provides insights into the trade-offs between model size, providers, and training strategies when applying post-training based LLM code generation optimization techniques in real-world settings.

\subsection{RQ1: Chained system vs. Individual techniques}
RQ1 investigates the impact of combining two widely adopted post-training techniques—multi-agent collaboration (ACT) and run-time execution information-based debugging (Debug)—on the functional accuracy of large language models (LLMs). We seek to determine how the combined application of these techniques compares to their individual application in enhancing functional accuracy. A series of experiments and statistical analyses were conducted to determine whether the combined application of ACT and Debug (ACT + Debug) offered statistically significant improvements in functional accuracy over ACT or Debug alone.

To answer RQ1, we tested 19 LLMs under three conditions: (1) the combined approach (ACT + Debug), (2) ACT alone, and (3) Debug alone. Comprehensive model wise results are provided in \hyperref[table2]{Table 2}. The functional accuracy for each model was evaluated under the three conditions using the HumanEval benchmark, offering a standardized performance measure, while the results of HumanEval+ are reserved for later assessment of code rigorousness. The mean accuracy for the combined approach was found to be 64.82\%, while ACT alone achieved a mean accuracy of 57.16\%, and Debug alone achieved a mean accuracy of 63.86\%. These observed differences served as the basis for hypothesis testing to assess the statistical significance of the improvements.

Two sets of hypotheses were formulated to guide the analysis. The first compared the combined approach (\(\text{ACT + Debug}\)) with ACT alone. The null hypothesis (\(H_{0,1}\)) posited that the mean functional accuracy of the combined approach would be less than or equal to that of ACT alone (\(\mu_{\mathrm{ACT+Debug}} \leq \mu_{\mathrm{ACT}}\)). Conversely, the alternative hypothesis (\(H_{1,1}\)) suggested that the combined approach would outperform ACT alone (\(\mu_{\mathrm{ACT+Debug}} > \mu_{\mathrm{ACT}}\)).

The second set of hypotheses compared the combined approach with Debug alone. The null hypothesis (\(H_{0,2}\)) posited that the mean functional accuracy of the combined approach would be less than or equal to that of Debug alone (\(\mu_{\mathrm{ACT+Debug}} \leq \mu_{\mathrm{Debug}}\)). The alternative hypothesis (\(H_{1,2}\)) proposed that the combined approach would significantly outperform Debug alone (\(\mu_{\mathrm{ACT+Debug}} > \mu_{\mathrm{Debug}}\)).

To test these hypotheses, a one-tailed paired t-test was employed. This method was selected due to the within-subject design of the study, in which each model was evaluated under all three experimental conditions (ACT, Debug, and ACT + Debug), ensuring consistency in testing across conditions. The t-statistic was calculated using the standard formula for paired samples shows as equation (1).

\begin{equation}
t = \frac{\bar{x} - \mu_0}{\frac{s}{\sqrt{n}}}
\label{eq}
\end{equation}
\vspace{0.05cm}

We take degrees of freedom (\(df\)) equal to \(n - 1 = 18\) and a significance level (\(\alpha\)) set at 0.15. The choice of a higher \(\alpha\) (\(\alpha = 0.15\)) aligns with our research objective of detecting even small performance gains to enhance the development and optimization of LLM techniques and guide deployment decisions in automated systems where even slight improvements in code generation can be beneficial. Slight gains can translate into improved efficiency, fewer errors, and higher productivity for developers. Additionally, since LLM-generated code serves as the foundation for future tasks, even marginal improvements can compound over time, amplifying the overall benefits. A higher alpha helps detect these subtle but meaningful improvements, ensuring that promising advancements are not overlooked due to an overly strict significance threshold. Moreover, code generation often occurs in interactive settings, where developers can verify and refine the output, further reducing the consequences of Type I errors. 

\subsubsection{Comparing the chained approach (ACT + Debugger) with ACT}

\begin{figure}[!htb]
    \centering
    \includegraphics[width=3.2in,height=3.0in,clip,keepaspectratio]{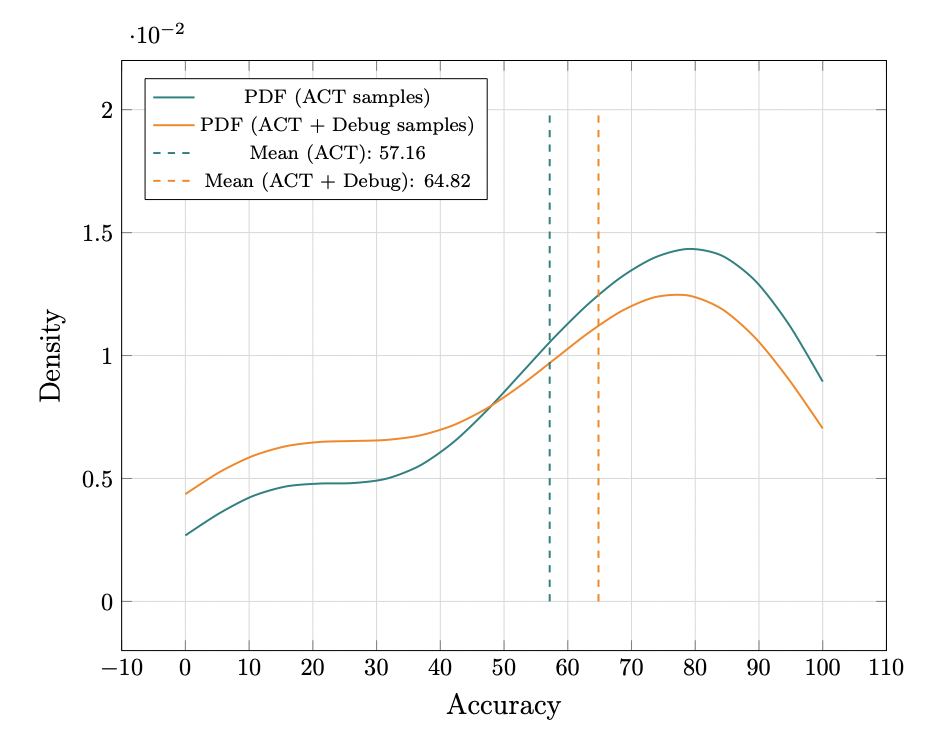}    \caption{Comparison of probability density functions for ACT and ACT+Debug samples on the HumanEval dataset.}
    \label{figure2}
\end{figure}

The results of the t-tests revealed a significant improvement in functional accuracy when comparing the chained approach to ACT alone (\(H_{0,1}\) rejected). Therefore, at the 15\% significance level, there is strong evidence to suggest that the chained framework (ACT + Debug) improves the average functional accuracy relative to ACT alone. The differences between the means of the two approaches are evident in the probability density functions of the ACT and ACT+Debug samples, as illustrated in \hyperref[figure2]{Figure 2}. The benefits of incorporating a debugger into the ACT approach is also evident in the results table (\hyperref[table2]{Table 2}), where nearly all LLMs exhibit improved accuracy in the HumanEval benchmark when the debugger is added.

\subsubsection{Comparing the chained approach (ACT + Debugger) with Debugger}

\begin{figure}[!htb]
    \centering
    \includegraphics[width=3.2in,height=3.0in,clip,keepaspectratio]{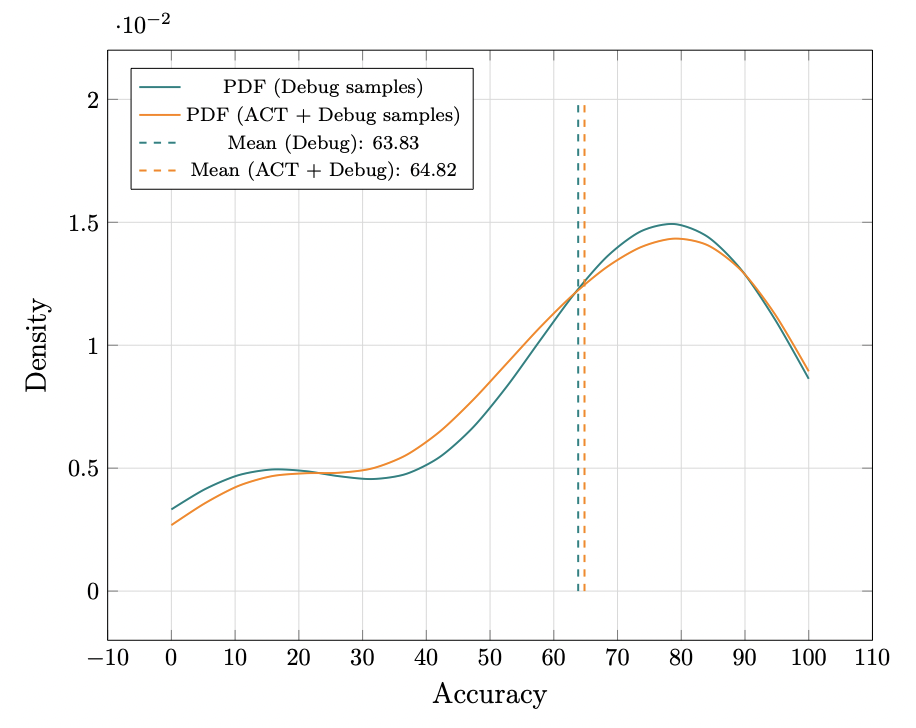}    \caption{Comparison of probability density functions for Debug and ACT+Debug samples on the HumanEval dataset.}
    \label{Figure 3}
\end{figure}

The improvement in accuracy when comparing the chained approach to Debug alone was 0.96\%. This difference was not statistically significant (\(H_{0,2}\) not rejected) at a significance level (\(\alpha\)) set at 0.15. The differences between the means of Debug and ACT+Debug can be observed in the probability density functions of the Debug and ACT+Debug samples, as illustrated in \hyperref[Figure 3]{Figure 3}.

These findings suggest that chaining ACT and Debug provides a significant improvement in functional accuracy when compared to ACT alone. However, the enhancement over Debug alone, while present, did not reach statistical significance under the conditions tested. This implies that Debug, as a standalone technique, may already capture much of the improvement facilitated by the combined approach. This is likely because, by breaking programs into smaller blocks and monitoring the values of intermediate variables during execution, Debug helps LLMs concentrate on more manageable sections of code, ensuring their accuracy and identifying errors more efficiently. The detailed contextual feedback provided during debugging—such as how intermediate calculations match expected results—enhances the LLM’s ability to iteratively improve the code. This effectiveness reduces the scope for complementary techniques such as ACT to deliver significant additional advantages. However, there are instances where the chained approach may still offer benefits, which we explore in the next section. 

% \begin{table*}[!htb]
%     % \centering
%     \includegraphics[width=10.8\textwidth,height=6.0in,clip,keepaspectratio]{table a:2 - results table.png} % Adjust the width as needed
%             \caption{\parbox{\textwidth}{\justifying Comparisons of Pass@1 accuracy scores for 19 language models on the HumanEval and HumanEval+ benchmarks are shown across six approaches: Basic, AC, ACT, Debugger, AC+Debugger, and ACT+Debugger. The last four columns display performance improvements (\(\Delta\)) for AC+Debugger and ACT+Debugger compared to ACT and Debugger, quantifying the value added by combining these approaches. Red highlights indicate a decrease in performance, while green highlights represent an increase in performance. Accuracy scores and \(\Delta\) are measured in percentages.}}

%     \label{table2}
% \end{table*}

\begin{table*}[!htb]
    \centering
    \includegraphics[width=10.8\textwidth,height=6.0in,clip,keepaspectratio]{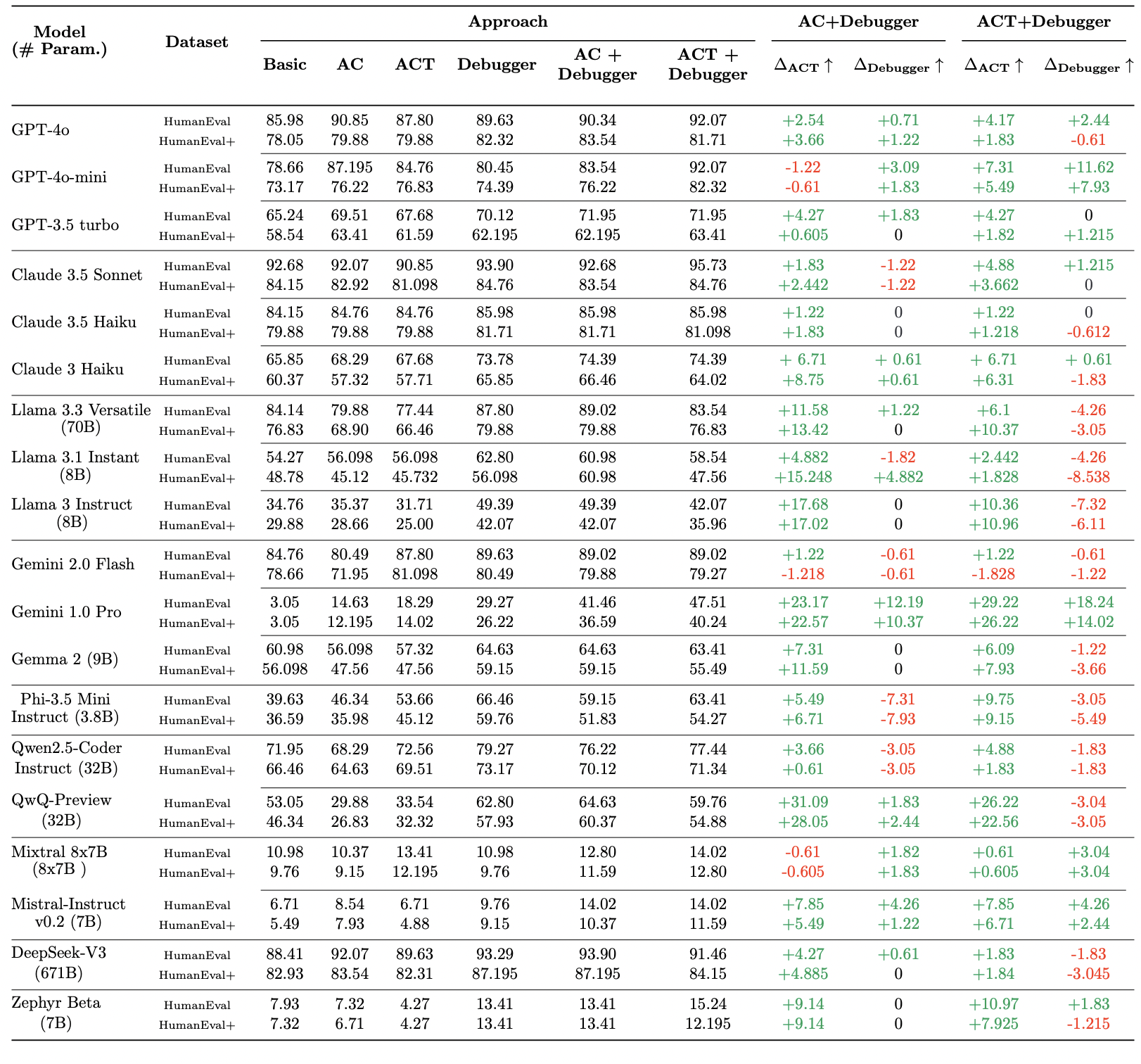} % Adjust the width as needed
    \caption{\justifying Comparisons of Pass@1 accuracy scores for 19 language models on the HumanEval and HumanEval+ benchmarks are shown across six approaches: Basic, AC, ACT, Debugger, AC+Debugger, and ACT+Debugger. The last four columns display performance improvements for AC+Debugger and ACT+Debugger compared to ACT and Debugger, quantifying the value added by combining these approaches. Red highlights indicate a decrease in performance, while green highlights represent an increase in performance. Accuracy scores and performance improvements are measured in percentages.}
    \label{table2}
\end{table*}

\subsubsection{Further exploration of the chained approach (ACT + Debugger) vs. Debugger on HumanEval and HumanEval+}

In this section, we delve deeper into the comparison between the chained approaches (ACT + Debugger, AC + Debugger) and the Debugger-Only approach, highlighting key observations and exploring scenarios where the chained approach may yield further benefits.

\begin{itemize}
\item Looking closely st the results table (\hyperref[table2]{Table 2}) we can observe that the addition of ACT to debugging shows varied benefits depending on the model. For frontier intelligent closed-source models like GPT-4o and Claude 3.5 Sonnet, the combination provides some improvements on the HumanEval benchmark (+2.44 and +1.215), but not HumanEval+ (-0.61 and 0), suggesting that ACT enhances the effectiveness of debugging in less rigourous datasets. For some smaller models like GPT-4o-mini, Mixtral 8x7B and Mistral-Instruct v0.2 (7B), ACT plays a more significant role in enhancing performance in both HumanEval and HumanEval+ (+11.62 and +7.93, +3.04 and +3.04, +4.26 and +2.44). On the other hand, for some models, such as GPT-3.5 turbo, Claude 3.5 Haiku or Claude 3 Haiku, the impact of ACT is minimal or nonexistent. Thus, we conclude that the decision to use ACT alongside debugging is highly dependent on the specific model and its internal mechanisms.

\item OpenAI and Anthropic have established themselves as leading AI research companies, making significant contributions to artificial intelligence through the development of innovative, closed-source large language models. With a primary focus on the advancement of LLM technology, both companies benefit from a large talent pool and  substantial funding from various investors. Additionally, given the consistently high performance of their models, organizations with the financial resources to utilize their services are likely to prioritize these providers when selecting an LLM for large-scale code generation. This motivated us to explore their models in greater depth. As visualized in \hyperref[Figure 4]{Figure 4}, we can observe that in the three OpenAI models we tested, the combinatorial approaches consistently improve performance over the Debugger approach. The ACT + Debugger approach achieves the highest accuracy among all OpenAI models, attaining 92.07\% on GPT-4o-mini and GPT-4o for the HumanEval dataset. This represents a notable improvement over the Debugging approach alone, which achieved 80.4/5\% with GPT-4o-mini and 89.63\% with GPT-4o. The AC + Debugger approach also demonstrates significant improvements over the Debugger method, achieving, for example, 83.54\% using GPT-4o-mini in HumanEval, representing a +3.09\% increase over the Debugger only approach. For the more challenging HumanEval+ dataset, all approaches exhibit performance degradation while preserving similar relative performance trends. The ACT + Debugger approach consistently outperforms other methods, with the exception of GPT-4o, where the AC + Debugger approach surpasses the Debugger approach, while the ACT + Debugger approach does not. This may be attributed to the agent collaboration generating code that was particularly challenging to debug during this run. These findings indicate that OpenAI models generally exhibit stable performance improvements when employing a combined approach.

\begin{figure}[!htb]
    \centering
    \includegraphics[width=3.2in,height=3.0in,clip,keepaspectratio]{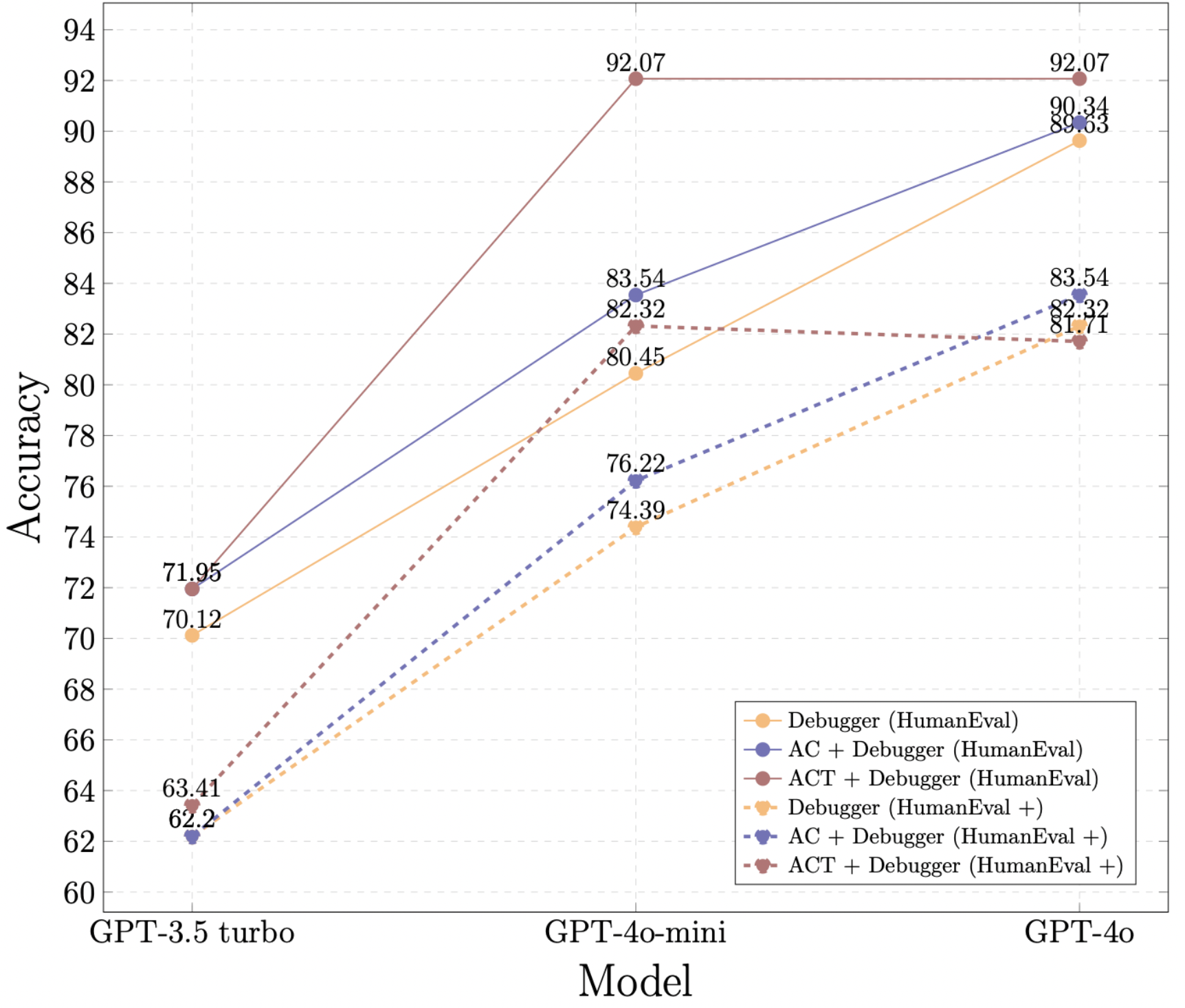}    \caption{\justifying Performance comparison of three OpenAI models on HumanEval and HumanEval+ datasets using three different approaches. The rightmost model on the x-axis has the highest token cost, while the leftmost is the most cost-efficient.}
    \label{Figure 4}
\end{figure}

\hyperref[Figure 5]{Figure 5} visualizes the results of the three Anthropic models tested. In particular, for Anthropic Claude 3.5 Sonnet, the combined ACT+Debugger approach yields significant improvements, achieving a peak performance of 95.73\% on HumanEval, a +1.83\% increase from the Debugger approach. However, the smaller Claude models from the Haiku family exhibit limited responsiveness to the combinatorial approaches, indicating that larger models from Anthropic are better suited for such enhancement strategies.

\begin{figure}[!htb]
    \centering
    \includegraphics[width=3.2in,height=3.0in,clip,keepaspectratio]{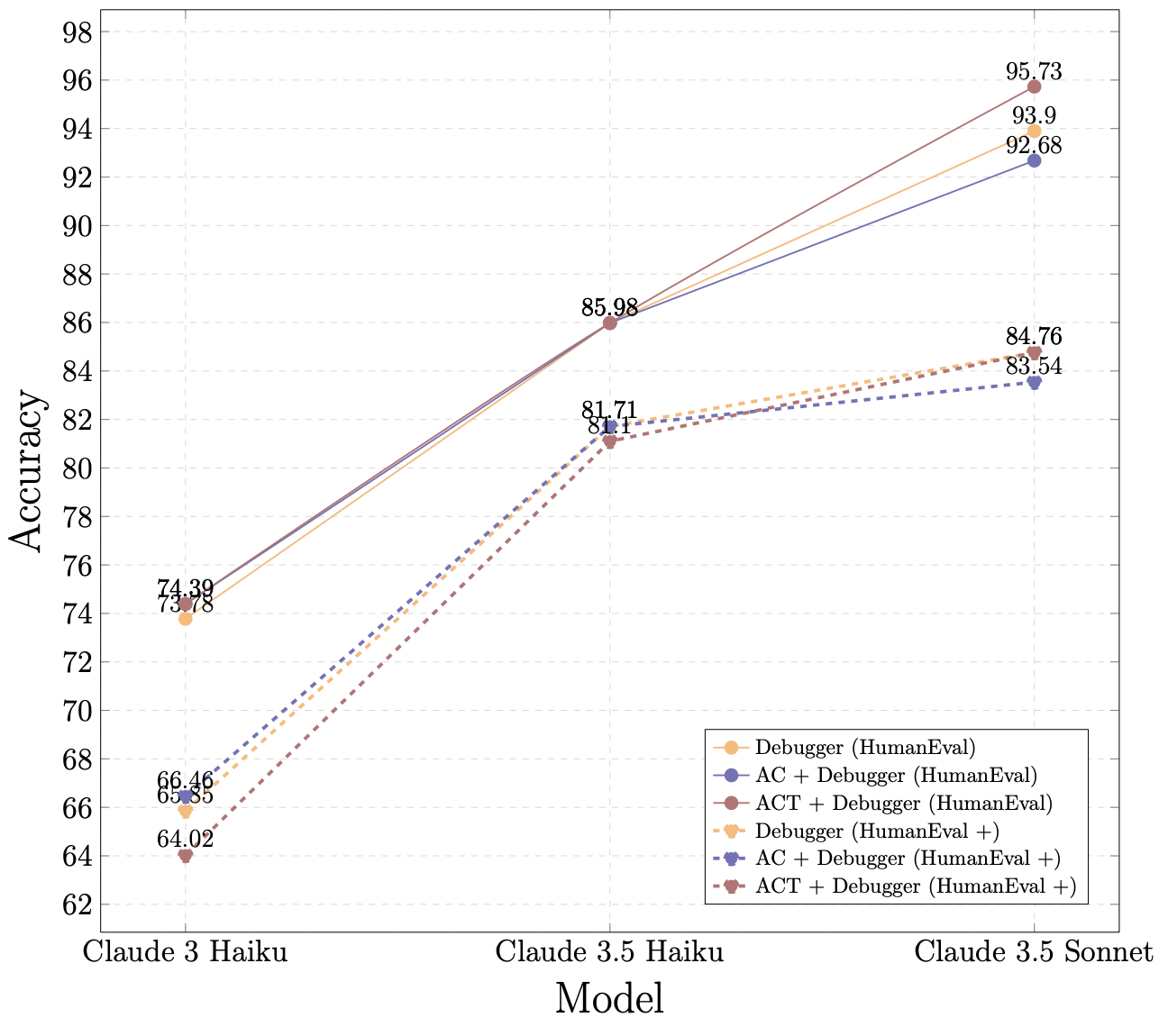}    \caption{\justifying Performance comparison of three Anthropic Claude models on HumanEval and HumanEval+ datasets using three different approaches. The rightmost model on the x-axis has the highest token cost, while the leftmost is the most cost-efficient.}
    \label{Figure 5}
\end{figure}

\item When considering open-sourced models such as Llama, Gemini, DeepSeek, and Mistral, the impact of ACT varies further. Llama, Gemini, and DeepSeek models did not show benefits from adding ACT to debugging, suggesting that the debugging process alone is sufficient for these models. However, Gemini 1.0 Pro stands out as an exception, showing a significant improvement when ACT is combined with debugging. This highlights that even within a group of models that generally do not benefit from ACT, certain variants may still experience a large performance boost. In contrast, Zephyr appears to have mixed reaction to the addition of ACT, with results from HumanEval showing improvement (+1.83) while a decline in HumanEval+ (-1.215). Table a further reveals that certain models, such as Llama 3.3 Versatile (70B), Llama 3.1 Instant, Llama 3 Instruct, Gemma 2 (9B), QwQ-Preview (32B) and DeepSeek-V3 (671B), demonstrate improved performance when agentic complexity is reduced. This improvement is evident as accuracy increases when using ACDebugger instead of ACTDebugger, suggesting that these models might not be optimal at handling agentic complexity efficiently.

\begin{figure}[!htb]
    \centering
    \includegraphics[width=3.2in,height=3.0in,clip,keepaspectratio]{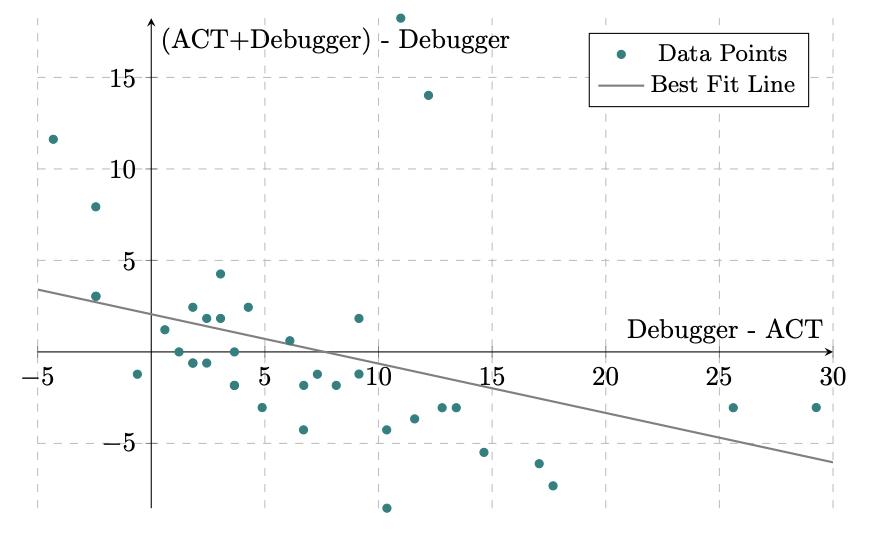}    \caption{Corellation between Debugger-ACT gap and ACT+Debug improvement in accuracy.}
    \label{Figure 6}
\end{figure}

\begin{figure}[!htb]
    \centering
    \includegraphics[width=3.2in,height=3.0in,clip,keepaspectratio]{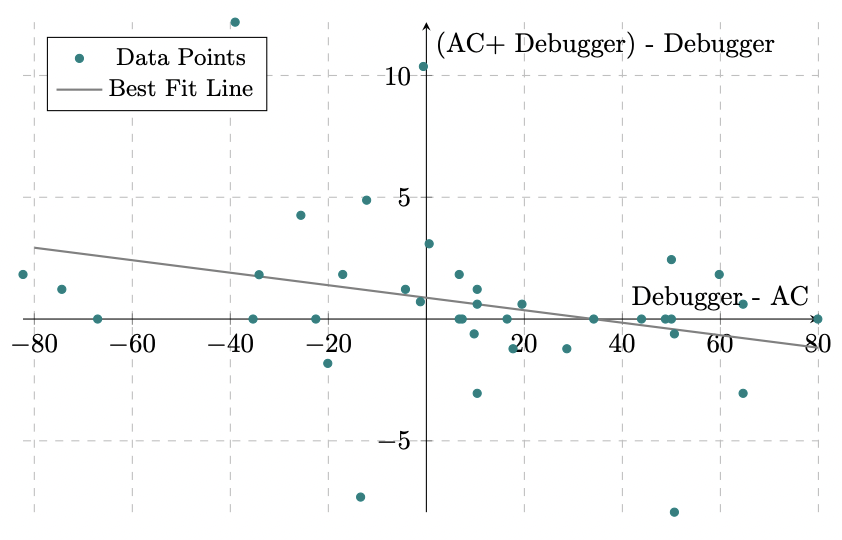}    \caption{Corellation between Debugger-AC gap and AC+Debug improvement in accuracy.}
    \label{Figure 7}
\end{figure}

\item In our analysis of model performance with ACT+Debug or AC+Debug versus Debug alone, we identified a clear pattern, which is visually represented in \hyperref[Figure 6]{Figure 6} and \hyperref[Figure 7]{Figure 7}. The relationship between the accuracy performance of the ACT approach and the Debugging approach, and their combination can be analyzed using \hyperref[Figure 6]{Figure 6}. This scatter plot compares the difference in accuracy between the chained approach (ACT+Debugger) and Debugger (y-axis) against the difference between Debugger and ACT (x-axis) for 38 data points across 2 datasets and 19 models. The trend captured by the best-fit line reveals an inverse correlation: the larger the gap between Debugger and ACT performance for a given model, the less benefit is observed from combining ACT with Debugging. In other words, when Debugging alone significantly outperforms ACT (as indicated by higher values on the x-axis), the incremental benefit of adding ACT to Debugging diminishes. Conversely, when ACT performs closer to Debugger (smaller values on the x-axis), the combination of the two methods tends to yield greater performance improvements (higher values on the y-axis). This aligns with the insights from \hyperref[table2]{Table 2}, emphasizing that ACT + Debugger is most effective when ACT performance is not vastly overshadowed by Debugging.

Building on the analysis, \hyperref[Figure 7]{Figure 7} reveals a trend similar to that observed in \hyperref[Figure 6]{Figure 6}. In this case, the comparison focuses on the accuracy difference between the chained approach (AC+Debugger) and Debugger, plotted against the difference between Debugger and AC. By replacing ACT with AC, the agentic complexity is effectively reduced. The plot includes 38 data points derived from 2 datasets and 19 models. Despite this simplification, the pattern remains consistent: as the performance gap between AC and Debugger widens, the advantage of integrating AC with Debugger diminishes.

Observation from \hyperref[Figure 6]{Figure 6} and \hyperref[Figure 7]{Figure 7} solidifies a hypothesis that the relationship between AC/ACT and Debugger accuracy governs whether adding AC/ACT to the Debugger only approach will yeild benefit as the performance improvement from combining the two methods is more pronounced when AC/ACT performance is closer to Debugger performance and diminishes as Debugger performance surpasses AC/ACT significantly. This trend emphasizes that the added benefit of chaining approaches becomes more apparent for models or cases where Debugger alone is less effective. Thus, for better chances of improved performance, it is advisable to focus on models where AC/ACT and Debugger perform equally well, as these models are more likely to benefit from the chained approach (ACT+Debugger). 

\end{itemize}

Ultimately, the effectiveness of ACT in combination with Debugging depends on the specific model, how well it handles agentic complexities, and how well it can understand the context provided by the debugging approach. For OpenAI models like GPT-4o, GPT-4o-mini and the series from Mistral, the combination of ACT and debugging can lead to noticeable gains, whereas for open-source models such as Llama, DeepSeek, and certain variants of Gemini, the benefit may be minimal or even absent. However, the substantial improvement in Gemini 1.0 Pro underscores that there is no one-size-fits-all rule, and each model’s specific performance characteristics and requirements must be considered when deciding whether ACT should be integrated with debugging.

\subsection{RQ2: Segment-wise performance}
% \vspace{0.5cm}
\subsubsection{Segment Details}

\begin{figure}[!htb]
    \centering
    \includegraphics[width=3.2in,height=3.0in,clip,keepaspectratio]{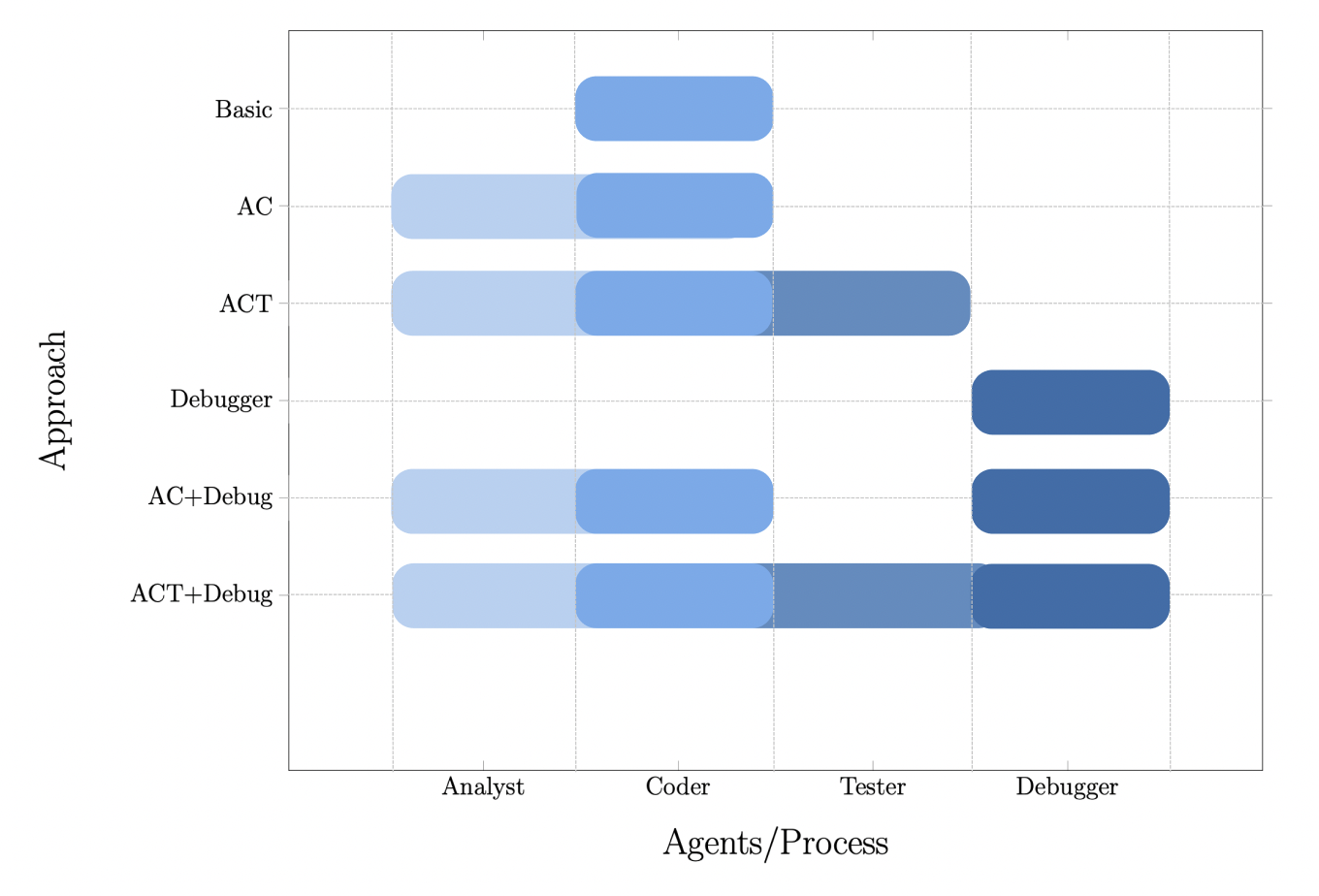}    \caption{Composition of the different segments explored in this paper.}
    \label{Figure 8}
\end{figure}

To investigate the influence of different components on the functional accuracy of the generated code, we decomposed our chained approach into several distinct segments. The rationale for segmenting is based on the assumption that the performance of each segment is influenced by the interactions among its constituent elements. The complete chain consists of two primary modules: multi-agent collaboration (ACT) and debugger, as can be observed in \hyperref[figure1]{Figure 1}. These modules can be configured into various pathways or segments, each representing a unique approach to code generation. From this point on, we will use the terms 'segments' and 'approaches' interchangeably. We examined six distinct approaches—Basic, AC, ACT, Debugger, AC+Debug, and ACT+Debug—to understand their individual contributions to accuracy. The composition of each approach can be observed in \hyperref[Figure 8]{Figure 8}.

The most basic approach employs a single coder agent, provided with role instructions and general guidelines to generate code directly from requirements. Building upon this, the Analyst + Coder (AC) approach introduces a requirement analyst agent who first creates a high-level plan before passing it to the coder agent to generate the corresponding code. Both agents receive role-specific prompting and basic instructions, although few-shot examples are not utilized in this experiment.

The ACT approach further enhances the process by incorporating a tester agent alongside the analyst and coder. It is one of the foundational techniques discussed in the methodology section. The tester's role involves conducting comprehensive code reviews and providing feedback to improve code quality. The tester can engage in up to 3 iterations (retriesCT) of interaction with the coder to resolve issues. If problems persist after three attempts, the most recent version of the code is considered final. A retriesCT of 3 is a reduction from the 4 iteration limit specified in the original Self-Collaboration paper \citep{21}. This slight adjustment was made to enhance efficiency in token generation, reduce latency, and ensure consistency across experiments, as later on we chain approaches that introduce additional LLM calls and complexities.

The Debugger approach, another core technique outlined in the methodology section, divides generated programs into basic blocks and tracks the values of intermediate variables during execution. This provides detailed context about the program, allowing the LLM to effectively identify and correct errors. Within this approach, the debug cycle (retriesD) is allowed to repeat up to 10 times. This choice aligns with the original LDB paper \citep{23}, which set 10 as the maximum number of debugging iterations.

The ACT + Debug approach represents the full chain implementation as outlined in our methodology, while the AC + Debug approach excludes the tester, instead routing the code directly to the debugging process if execution against visible test cases reveals issues. The debug cycle (retriesD) can be repeated up to four times when using the AC + Debug or ACT + Debug approaches. This decision was made to save time and reduce costs, as these combined approaches require more LLM calls than the Debugger approach. Tester engagement (retriesCT) is capped at 3 interactions, consistent with the ACT approach, to maintain consistency across experiments.

By dividing our chained system into distinct segments or approaches, our experimental design enables us to isolate and evaluate the contribution of each segment to the overall functional accuracy. All agent prompts, including those used in the debugging process, are can be found in our GitHub repository \textsuperscript{\hyperref[fn:github]{\ref{fn:github}}}.

\vspace{0.5cm}
\subsubsection{Segment Performance}

\begin{figure*}[!htb]
    \centering
    \includegraphics[width=\textwidth,clip,keepaspectratio]{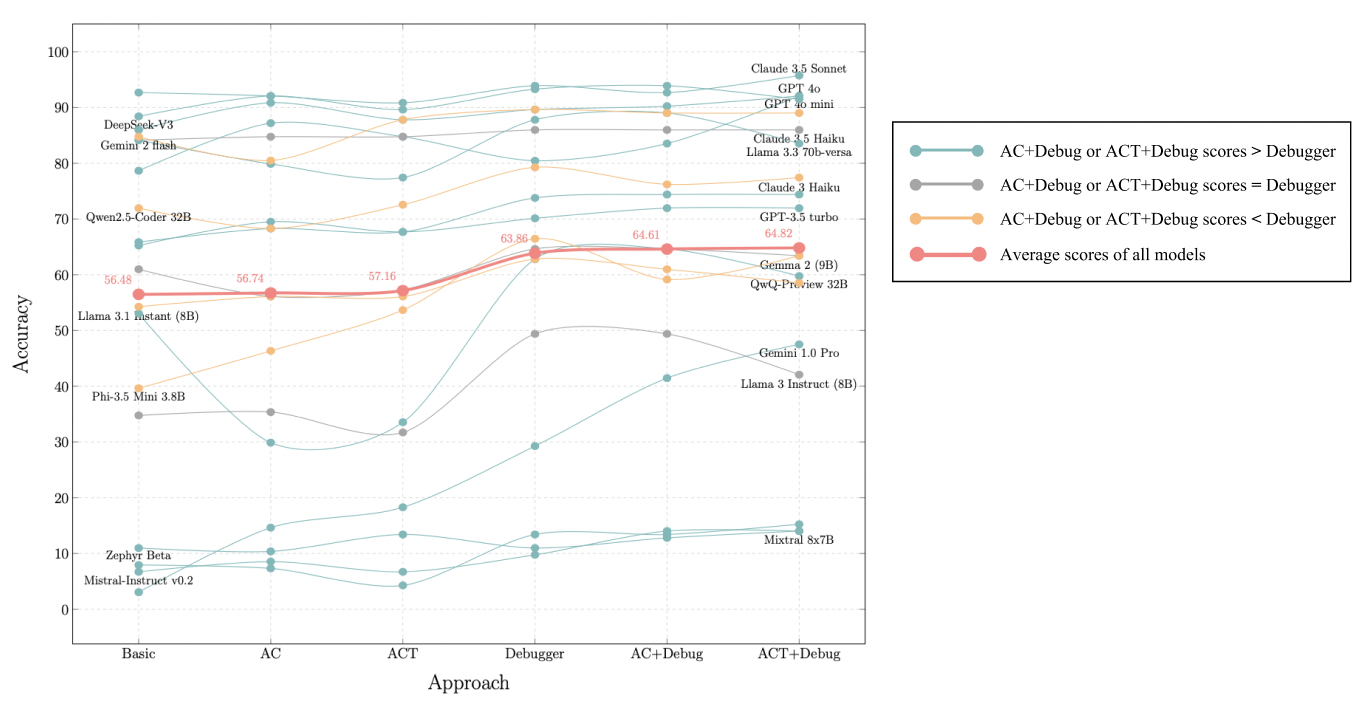}
    \caption{\justifying Pass@1 accuracy scores of 19 Models tested with 6 different approaches on the HumanEval benchmark. Green lines indicate models where AC+Debug or ACT+Debug scores surpass those of the Debugger, gray lines represent cases where the scores are equal, and orange lines denote instances where the scores are lower. The red line represents the average scores of all models. Accuracy scores are measured in percentages.}
    \label{Figure 9}
\end{figure*}

\begin{figure*}[!htb]
    \centering
    \includegraphics[width=\textwidth,clip,keepaspectratio]{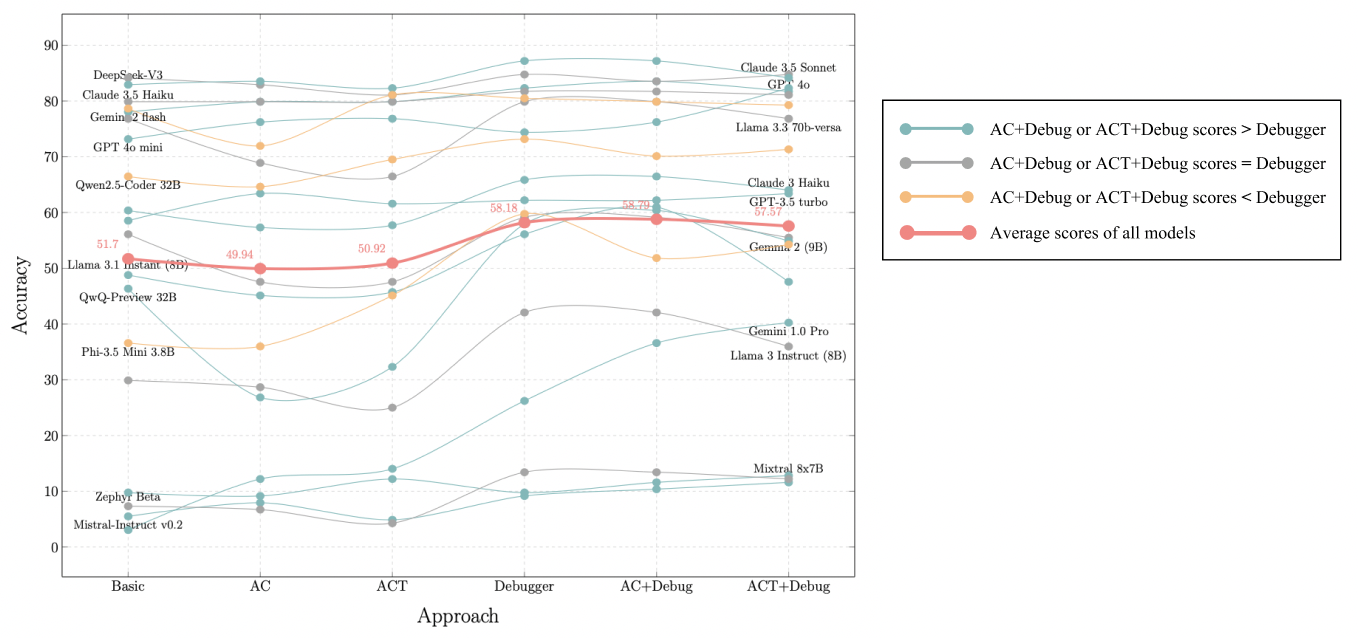}
    \caption{\justifying Pass@1 accuracy score of 19 Models tested with 6 different approaches on the HumanEval+ benchmark. Green lines indicate models where AC+Debug or ACT+Debug scores surpass those of the Debugger, gray lines represent cases where the scores are equal, and orange lines denote instances where the scores are lower. The red line represents the average scores of all models. Accuracy scores are measured in percentages.}
    \label{Figure 10}
\end{figure*}

The performance of different approaches can be observed in the results table (\hyperref[table2]{Table 2}). To better illustrate the increase in performance from approach to approach, we plot \hyperref[Figure 9]{Figure 9} (HumanEval) and \hyperref[Figure 10] {Figure 10} (HumanEval+). Given that the debugger and agentic interactions with the debugger-based approaches produce the highest accuracy, with small differences in performance, we plot the models as follows: in green, where AC+Debug or ACT+Debug scores exceed the Debugger’s; in gray, where scores are equal; and in orange, where scores are lower. This visualization highlights models that may benefit from incorporating agentic interactions into the debugging process.

\hyperref[Figure 9] {Figure 9} and \hyperref[Figure 10] {Figure 10} illustrates the evolution of functional accuracy for the code generated by 19 tested models across 6 approaches on the HumanEval and HumanEval+ datasets. The red line denotes the average performance across all models. The mean accuracy in both cases improves as the process progresses from the Basic approach to more sophisticated methods, such as ACT + Debug, demonstrating the value of incorporating additional components like analysis, testing, and debugging. Interestingly, the best-performing approach on the HumanEval+ dataset is not the most complex, but rather the AC+Debug approach. This observation indicates that models generally perform better on datasets with rigorous evaluation criteria when the agentic complexity is reduced and the process remains straightforward. The Basic approach, involving a single coder agent, shows the lowest mean accuracy on the HumanEval dataset but not on HumanEval+, further highlighting this observation, that lower agentic interaction yields more rigorous code.

An interesting pattern is observed in the case of QwQ-Preview (32B), where both the AC and the ACT approaches perform significantly worse than the basic approach in both datasets. This performance decline may be attributed to the interaction between the analyst and the coder, which may have introduced misleading information for a substantial number of problems. Consequently, the confusion introduced at this stage probably propagated through the subsequent steps, ultimately resulting in incorrect outputs. For several models, including OpenAI models, Claude 3.5 Sonnet, Llama models, and DeepSeek-V3, the ACT approach performed worse than the AC segment. This indicates that while adding a tester is often beneficial, it can occasionally introduce complexities or inefficiencies that negatively impact performance.

The Debugger approach, even when applied independently, achieves significant accuracy improvements over ACT, with the mean accuracy increasing by 6.7\% in HumanEval and 7.36\% in HumanEval+. The Debugger approach excels due to its ability to identify and rectify errors in a systematic manner. Unlike methods that rely solely on generating code, the Debugger actively executes the generated code against test cases, detecting failures, providing rich context about those failures, and iteratively refining the output. 
 
Integrating debugging with multi-agent collaboration achieves the higher functional accuracy. In the AC + Debug approach, the interaction between the analyst and coder generates code that undergoes debugging if visible test cases fail. This approach results in slight mean accuracy improvements of 0.75\% in HumanEval and 0.61\% in HumanEval+.

The complete chain, represented by the ACT + Debug segment, achieves the highest overall accuracy in the HumanEval dataset, showing an improvement of 0.21\% over the AC + Debug approach. This shows the effectiveness of integrating all components to maximize the quality and functionality of the generated code. However, consistent with the earlier finding that reduced agentic interaction leads to more rigorous code, the AC + Debug approach achieves the highest mean accuracy on the HumanEval+ dataset. The complete chain (ACT + Debug) performs 1.22\% worse, highlighting that added complexity can introduce errors or inconsistencies, especially in datasets with strict evaluation criteria such as HumanEval+.

\subsection{RQ3: Code Rigorousness}
% \vspace{0.5cm}

The rigor of the code generated by each segmented approach was evaluated by examining the accuracy drop between the HumanEval and HumanEval+ datasets, the latter being a significantly more stringent evaluation framework with 80 times more tests. Highly rigorous code ensures software is reliable, maintainable, and resilient in demanding environments, making it crucial for LLM-generated code to be rigorous. 

\subsubsection{Approach Consideration}

\begin{figure}[!htb]
    \centering    \includegraphics[width=3.2in,height=3.0in,clip,keepaspectratio]{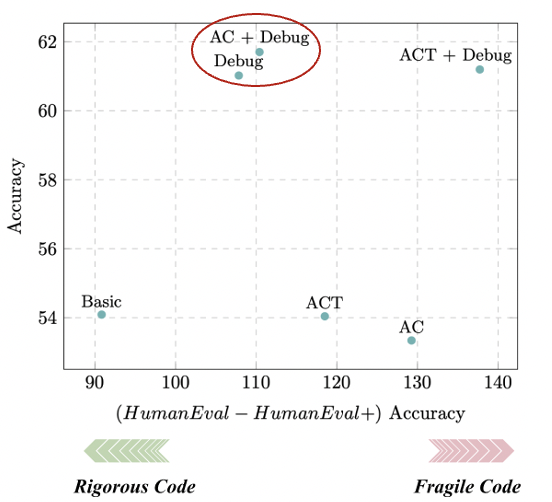}    \caption{\justifying Mean Pass@1 accuracy of 6 approaches across both datasets, compared to the average accuracy drop when the generated code was evaluated using 80 times more tests (HumanEval+). The average accuracy drop is the sum of Pass@1 accuracy differences for a given approach across two datasets using all LLMs.}
    \label{Figure 11}
\end{figure}

\hyperref[Figure 11] {Figure 11} illustrates the relationship between the mean accuracy in all six approaches and their performance degradation when subjected to expanded test coverage (HumanEval+). A smaller accuracy differential between HumanEval and HumanEval+ indicates that the code generated by the approach was highly rigorous. Our analysis revealed distinct trends in code rigor and functional accuracy across approaches. The Basic approach exhibited the highest rigor, with the smallest accuracy drop (90.83), highlighting its ability to maintain functionality under rigorous testing conditions despite producing comparatively lower overall accuracy. However, its simplicity and lack of advanced error correction mechanisms limit its utility in complex scenarios.

AC and ACT approaches demonstrated mean functional accuracy similar to the basic approach but exhibited substantial drops in robustness (129.27 and 118.51, respectively), indicating that increased agentic interaction introduces fragility. The Debugger Only approach achieved a compelling balance of high accuracy and robustness, with an accuracy drop of 107.84. By leveraging runtime execution information to directly address functional issues, debugging significantly enhances the generalization capability of generated code. Similarly, the AC + Debugger approach integrates planning with debugging, resulting in a slight decrease in robustness (110.41) compared to the Debugger Only approach, but achieving a 0.68\% increase in mean accuracy. This configuration represents a balanced approach, effectively mitigating the fragility observed in purely agent-driven methods, while achieving the highest accuracy.

Lastly, the ACT + Debugger approach, while achieving the highest accuracy, showed the highest drop in robustness (137.74). This suggests that the compounded complexity of multi-agent collaboration and iterative feedback loops reduces the ability to produce code resilient to diverse and stringent testing conditions. These findings further support the observations from RQ2, that reduced agentic interaction generally leads to more reliable code generation, emphasizing the importance of simplicity and runtime feedback mechanisms in achieving rigor under demanding evaluation scenarios.

\subsubsection{Model Consideration}

\begin{figure}[!htb]
    \centering
    \includegraphics[width=3.2in,height=3.0in,clip,keepaspectratio]{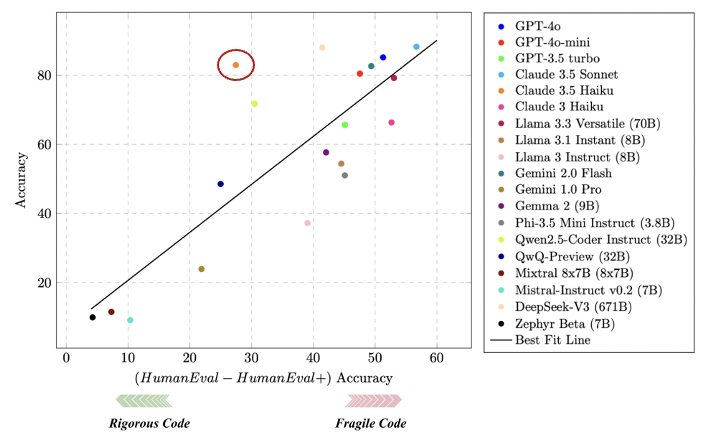}    \caption{\justifying Model specific mean Pass@1 accuracy score (of all six approaches across both datasets), compared to model specific average accuracy drop (across all approaches), when the generated code was evaluated using 80 times more tests (HumanEval+).}
    \label{Figure 12}
\end{figure}

\hyperref[Figure 12] {Figure 12}, compares the mean accuracy of each model across all six approaches to the corresponding drop in accuracy under the more stringent testing conditions of HumanEval+. This plot highlights notable differences in the rigor of the code produced by different LLMs. 

Advanced models such as Claude 3.5 Sonnet, GPT-4o, and DeepSeek-V3 consistently produce high-quality code. However, these top-performing models often exhibit significant accuracy drops under more rigorous testing conditions, as indicated by the positive correlation in the best-fit line. In paricular, Claude 3.5 Haiku stands out as an exception to this trend, demonstrating stable performance even with increased testing rigor. This suggests that Claude 3.5 Haiku offers the most favorable balance between accuracy and reliability, making it a strong candidate for generating rigorous code using any of the six approaches.

In contrast, smaller models such as Mistral-Instruct v0.2 (7B) and Zephyr Beta (7B) demonstrate both lower initial accuracy and smaller absolute performance drops. This minimal decrease in performance might be due to limited room for further deterioration in performance when subjected to more rigorous testing conditions. The line of best fit identifies a trade-off in current AI systems: models that achieve higher accuracy tend to be more susceptible to performance degradation under rigorous testing. This relationship holds true across different model sizes and architectures, from smaller models clustering in the lower left to larger models in the upper right, highlighting a core challenge in AI code generation: achieving and sustaining high performance under increasingly stringent evaluation conditions.

\begin{figure*}[!htbp]
    \centering
    \includegraphics[width=\textwidth,clip,keepaspectratio]{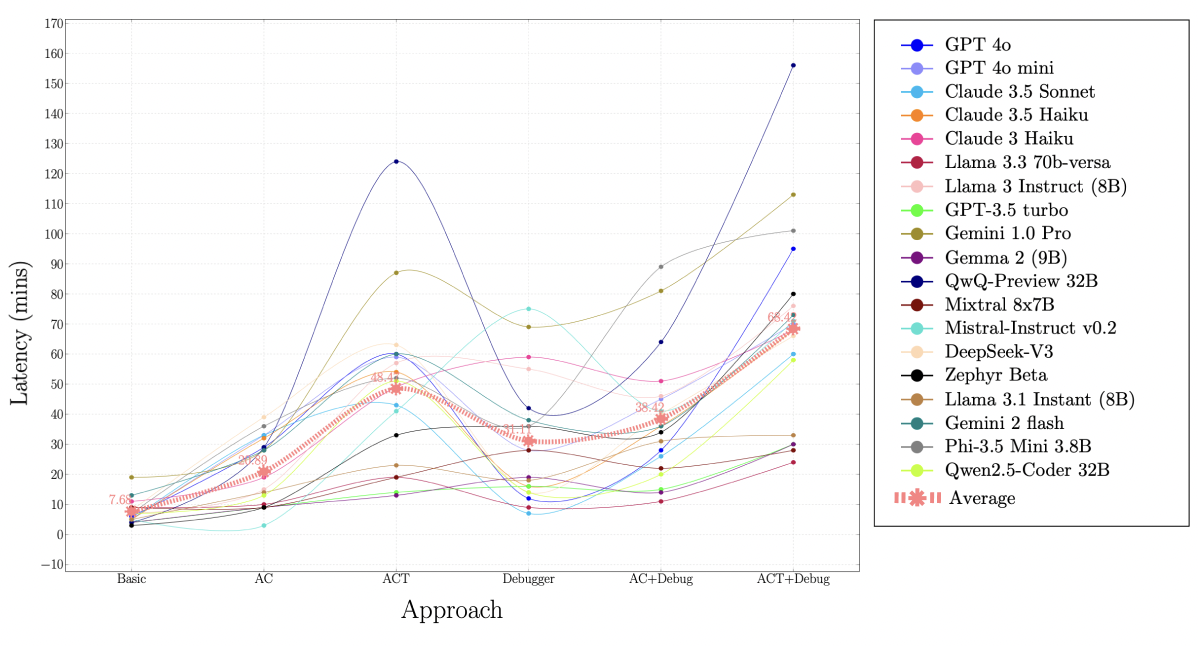}
    \caption{\centering Average time taken for each approach across 19 LLMs.}
    \label{Figure 13}
\end{figure*}

\subsection{RQ4: Latency Comparism}

The latency analysis reveals significant variations across the six approaches, with execution times ranging from approximately 8 to over 68 minutes. As shown in \hyperref[Figure 13]{Figure 13} and Table 3, the basic approach is the fastest, averaging 7.68 minutes, but achieves only modest accuracy (54.09\%). This represents the baseline trade-off - minimal time investment for acceptable but not optimal results.

The introduction of collaborative agents progressively increases latency. The AC approach(Analyst + Coder) nearly triples the execution time to 20.89 minutes without improving accuracy (53.34\%), suggesting that this configuration offers poor time-quality trade-offs. Introducing a tester agent (ACT) raises the average latency to 48.47 minutes while maintaining a similar accuracy of 54.04\%, suggesting that the increased time investment in agent-collaboration approaches does not result in proportional quality gains across the 19 LLMs we tested.

The standalone debugger approach emerges as a particularly efficient trade-off point, requiring 31.11 minutes on average while achieving a high average accuracy of 61.02\% across both datasets. This represents a 6.93\% average accuracy gain over the basic approach for a moderate increase in execution time (23.43 minutes). The AC + Debugger configuration attains the highest average accuracy (61.7\%) across both datasets while maintaining a reasonable execution time of 38.42 minutes, indicating an optimal balance between efficiency and code quality.

The full ACT + Debug chain, while theoretically comprehensive, demonstrates diminishing returns. It requires the longest execution time (68.42 minutes) but achieves slightly lower average accuracy (61.195\%) than the AC + Debugger approach, suggesting that the additional complexity and time investment of the complete chain may not be justified by the results.

Looking at individual model performance in \hyperref[Figure 13]{Figure 13}, we observe significant variance in latency across different LLMs, with some models showing particularly high execution times in the more complex configurations. This variance indicates that the choice of base model can significantly impact the practicality of different approaches, especially in time-sensitive applications.

In conclusion, the analysis reveals that, while more complex approaches generally require longer execution times, the relationship between time investment and quality improvements is not linear. The AC + Debugger configuration emerges as the optimal strategy, delivering significant accuracy improvements with moderate latency increases. For a concise overview, \hyperref[Table 3]{Table 3} provides a summary of the average latency and accuracy for each approach, evaluated across all 19 LLMs and both datasets. 

\begin{table}[H]
    \centering
    \includegraphics[width=3.2in,height=3.0in,clip,keepaspectratio]{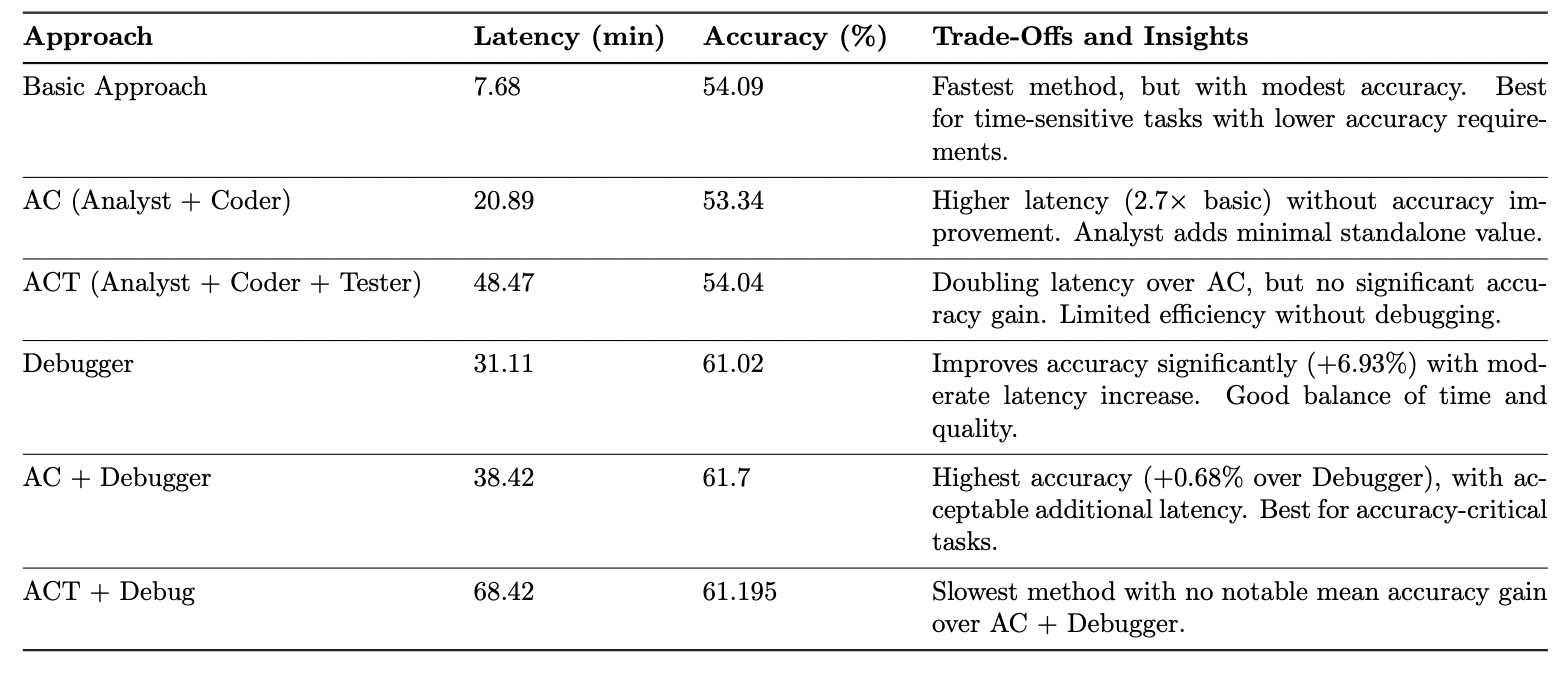} % Adjust the width as needed
    \caption{Summary of Latency, Accuracy, and Trade-Offs Across Approaches.}
    \label{Table 3}
\end{table}

\section{Conclusion}
In this paper, we investigate the effectiveness of combining multi-agent collaboration, as a process-based enhancement, with runtime execution information-based debugging, as a product-based enhancement, to improve LLM-driven code generation. Through extensive experiments on 19 LLMs across two benchmark datasets, our findings indicate that the debugging-based approach generally outperforms the agentic workflow. This may be attributed to the fact that agentic conversations tend to be hit-or-miss for many models, whereas a technique like debugging offers context-rich information that can benefit all models. However, the success of integrating both strategies hinges on the magnitude of the performance disparity between them. When debugging and agentic techniques show similar performance levels for a particular LLM, their combination yields better results. For models where debugging significantly outperforms the agentic approach, adding the latter introduces unnecessary complexity and diminishes benefits. Our findings further reveal that incorporating a simple two-agent workflow (Analyst-Coder) with debugging mechanisms results in a modest but meaningful 0.68\% improvement in mean accuracy, while preserving code rigor and maintaining similar latency across 19 LLMs, compared to using debugging alone. In contrast, more complex agentic configurations when used in combination with debugging often introduce higher latency without yielding substantial improvements in functional accuracy or code rigor. This suggests that across a wide range of LLMs, the most effective approach is to maintain a simple agentic workflow while leveraging a robust debugging mechanism. 

These insights have important implications for real-world AI applications, particularly in downstream tasks such as code completion, where even slight improvements in functional accuracy can yield significant benefits. Optimizing code generation requires not only enhancing functional accuracy, but also maintaining a balance between code rigor and generation latency, which is essential for leveraging pretrained LLMs effectively. By identifying the most effective post-training approaches and their optimal combinations, this research provides valuable insights for organizations seeking to deploy reliable, high-performance LLM-based systems, ultimately fostering the development of more robust and production-ready automated coding solutions.

% \bibitem{b1} G. O. Young, ``Synthetic structure of industrial plastics,'' in \emph{Plastics,} 2\textsuperscript{nd} ed., vol. 3, J. Peters, Ed. New York, NY, USA: McGraw-Hill, 1964, pp. 15--64.

% \vspace{6em}


\begin{thebibliography}{99}

\bibitem{1} OpenAI et al., “GPT-4o System Card,” Oct. 2024, [Online]. Available: \url{http://arxiv.org/abs/2410.21276}.
\bibitem{2} “The Claude 3 Model Family: Opus, Sonnet, Haiku Anthropic.” [Online]. Available: \url{https://docs.anthropic.com/}.
\bibitem{3} A. Grattafiori et al., “The Llama 3 Herd of Models,” Jul. 2024, [Online]. Available: \url{http://arxiv.org/abs/2407.21783}.
\bibitem{4} Gemini Team et al., “Gemini: A Family of Highly Capable Multimodal Models,” Dec. 2023, [Online]. Available: \url{http://arxiv.org/abs/2312.11805}.
\bibitem{5} Gemma Team et al., “Gemma 2: Improving Open Language Models at a Practical Size,” Jul. 2024, [Online]. Available: \url{http://arxiv.org/abs/2408.00118}.
\bibitem{6} H. Touvron et al., “LLaMA: Open and Efficient Foundation Language Models,” Feb. 2023, [Online]. Available: \url{http://arxiv.org/abs/2302.13971}.
\bibitem{7} M. Abdin et al., “Phi-3 Technical Report: A Highly Capable Language Model Locally on Your Phone,” Apr. 2024, [Online]. Available: \url{http://arxiv.org/abs/2404.14219}.
\bibitem{8} A. Yang et al., “Qwen2 Technical Report,” Jul. 2024, [Online]. Available: \url{http://arxiv.org/abs/2407.10671}.
\bibitem{9} A. Q. Jiang et al., “Mixtral of Experts,” Jan. 2024, [Online]. Available: \url{http://arxiv.org/abs/2401.04088}.
\bibitem{10} A. Q. Jiang et al., “Mistral 7B,” Oct. 2023, [Online]. Available: \url{http://arxiv.org/abs/2310.06825}.
\bibitem{11} DeepSeek-AI et al., “DeepSeek-V3 Technical Report,” Dec. 2024, [Online]. Available: \url{http://arxiv.org/abs/2412.19437}.
\bibitem{12} L. Tunstall et al., “Zephyr: Direct Distillation of LM Alignment,” Oct. 2023, [Online]. Available: \url{http://arxiv.org/abs/2310.16944}.
\bibitem{13} M. Chen et al., “Evaluating Large Language Models Trained on Code,” Jul. 2021, [Online]. Available: \url{http://arxiv.org/abs/2107.03374}.
\bibitem{14} H. Jin, L. Huang, H. Cai, J. Yan, B. Li, and H. Chen, “From LLMs to LLM-based Agents for Software Engineering: A Survey of Current, Challenges and Future,” Aug. 2024, [Online]. Available: \url{http://arxiv.org/abs/2408.02479}.
\bibitem{15} Z. Zhang et al., “Unifying the Perspectives of NLP and Software Engineering: A Survey on Language Models for Code,” Nov. 2023, [Online]. Available: \url{http://arxiv.org/abs/2311.07989}.
\bibitem{16} D. Huang, J. M. Zhang, M. Luck, Q. Bu, Y. Qing, and H. Cui, “AgentCoder: Multi-Agent-based Code Generation with Iterative Testing and Optimisation,” Dec. 2023, [Online]. Available: \url{http://arxiv.org/abs/2312.13010}.
\bibitem{17} Md. A. Islam, M. E. Ali, and M. R. Parvez, “MapCoder: Multi-Agent Code Generation for Competitive Problem Solving,” May 2024, [Online]. Available: \url{http://arxiv.org/abs/2405.11403}.
\bibitem{18} C. Qian et al., “Scaling Large-Language-Model-based Multi-Agent Collaboration,” Jun. 2024, [Online]. Available: \url{http://arxiv.org/abs/2406.07155}.
\bibitem{19} C. Qian et al., “ChatDev: Communicative Agents for Software Development,” Jul. 2023, [Online]. Available: \url{http://arxiv.org/abs/2307.07924}.
\bibitem{20} S. Hong et al., “MetaGPT: Meta Programming for A Multi-Agent Collaborative Framework,” Aug. 2023, [Online]. Available: \url{http://arxiv.org/abs/2308.00352}.
\bibitem{21} Y. Dong, X. Jiang, Z. Jin, and G. Li, “Self-collaboration Code Generation via ChatGPT,” Apr. 2023, [Online]. Available: \url{http://arxiv.org/abs/2304.07590}.
\bibitem{22} Y. Ding, M. J. Min, G. Kaiser, and B. Ray, “CYCLE: Learning to Self-Refine the Code Generation,” Proceedings of the ACM on Programming Languages, vol. 8, no. OOPSLA1, Apr. 2024, doi: 10.1145/3649825.
\bibitem{23} L. Zhong, Z. Wang, and J. Shang, “Debug like a Human: A Large Language Model Debugger via Verifying Runtime Execution Step-by-step,” Feb. 2024, [Online]. Available: \url{http://arxiv.org/abs/2402.16906}.
\bibitem{24} X. Hu, K. Kuang, J. Sun, H. Yang, and F. Wu, “Leveraging Print Debugging to Improve Code Generation in Large Language Models,” Jan. 2024, [Online]. Available: \url{http://arxiv.org/abs/2401.05319}.
\bibitem{25} H. Jin, Z. Sun, and H. Chen, “RGD: Multi-LLM Based Agent Debugger via Refinement and Generation Guidance,” Oct. 2024, [Online]. Available: \url{http://arxiv.org/abs/2410.01242}.
\bibitem{26} Y. Shi, S. Wang, C. Wan, and X. Gu, “From Code to Correctness: Closing the Last Mile of Code Generation with Hierarchical Debugging,” Oct. 2024, [Online]. Available: \url{http://arxiv.org/abs/2410.01215}.
\bibitem{27} X. Chen, M. Lin, N. Schärli, and D. Zhou, “Teaching Large Language Models to Self-Debug,” Apr. 2023, [Online]. Available: \url{http://arxiv.org/abs/2304.05128}.
\bibitem{28} F. Lin, D. J. Kim, Tse-Husn, and Chen, “SOEN-101: Code Generation by Emulating Software Process Models Using Large Language Model Agents,” Mar. 2024, [Online]. Available: \url{http://arxiv.org/abs/2403.15852}.
\bibitem{29} H. Ren, M. Zhan, Z. Wu, A. Zhou, J. Pan, and H. Li, “ReflectionCoder: Learning from Reflection Sequence for Enhanced One-off Code Generation,” May 2024, [Online]. Available: \url{http://arxiv.org/abs/2405.17057}.
\bibitem{30} X. Chen, M. Lin, N. Schärli, and D. Zhou, “Teaching Large Language Models to Self-Debug,” Apr. 2023, [Online]. Available: \url{http://arxiv.org/abs/2304.05128}.
\bibitem{31} J. Austin et al., “Program Synthesis with Large Language Models,” Aug. 2021, [Online]. Available: \url{http://arxiv.org/abs/2108.07732}.
\bibitem{32} J. Liu, C. Steven Xia, and Y. Wang Lingming Zhang, “Is Your Code Generated by ChatGPT Really Correct? Rigorous Evaluation of Large Language Models for Code Generation.” [Online]. Available: \url{https://github.com/evalplus/evalplus}.
\bibitem{33} B. Workshop et al., “BLOOM: A 176B-Parameter Open-Access Multilingual Language Model,” Nov. 2022, [Online]. Available: \url{http://arxiv.org/abs/2211.05100}.
\bibitem{34} A. Chowdhery et al., “PaLM: Scaling Language Modeling with Pathways,” 2023. [Online]. Available: \url{http://jmlr.org/papers/v24/22-1144.html}.
\bibitem{35} J. Bai et al., “Qwen Technical Report,” Sep. 2023, [Online]. Available: \url{http://arxiv.org/abs/2309.16609}.
\bibitem{36} D. Dai et al., “DeepSeekMoE: Towards Ultimate Expert Specialization in Mixture-of-Experts Language Models,” Jan. 2024, [Online]. Available: \url{http://arxiv.org/abs/2401.06066}.
\bibitem{37} B. Rozière et al., “Code Llama: Open Foundation Models for Code,” Aug. 2023, [Online]. Available: \url{http://arxiv.org/abs/2308.12950}.
\bibitem{38} Z. Luo et al., “WizardCoder: Empowering Code Large Language Models with Evol-Instruct,” Jun. 2023, [Online]. Available: \url{http://arxiv.org/abs/2306.08568}.
\bibitem{39} R. Li et al., “StarCoder: may the source be with you!,” May 2023, [Online]. Available: \url{http://arxiv.org/abs/2305.06161}.
\bibitem{40} Y. Wei, Z. Wang, J. Liu, Y. Ding, and L. Zhang, “Magicoder: Empowering Code Generation with OSS-Instruct,” Dec. 2023, [Online]. Available: \url{http://arxiv.org/abs/2312.02120}.
\bibitem{41} T. Zheng et al., “OpenCodeInterpreter: Integrating Code Generation with Execution and Refinement,” Feb. 2024, [Online]. Available: \url{http://arxiv.org/abs/2402.14658}.
\bibitem{42} A. Lozhkov et al., “StarCoder 2 and The Stack v2: The Next Generation,” Feb. 2024, [Online]. Available: \url{http://arxiv.org/abs/2402.19173}.
\bibitem{44} N. Shinn et al., “Reflexion: Language Agents with Verbal Reinforcement Learning,” Mar. 2023, [Online]. Available: \url{https://arxiv.org/abs/2303.11366}.

\end{thebibliography}
\end{document}